\documentclass[twocolumn,floatfix]{aastex63}
\usepackage{amsmath, amsfonts, amssymb,lmodern,mathrsfs,ulem,mathtools}
\usepackage{graphicx}
\usepackage{enumitem}
\usepackage{natbib}
\usepackage{xcolor}
\usepackage{makecell}
\usepackage{color}
\usepackage{hyperref}
\usepackage[utf8x]{inputenc}
\usepackage{tikz}

\newcommand{\Nv}{{\boldsymbol{N}}}
\newcommand{\Av}{{\boldsymbol{A}}}
\newcommand{\Cv}{{\boldsymbol{C}}}
\newcommand{\Ev}{{\boldsymbol{E}}}
\newcommand{\Sv}{\boldsymbol{\Sigma}}
\newcommand{\Bv}{\boldsymbol{B}}
\newcommand{\Fv}{\boldsymbol{F}}
\newcommand{\uv}{\boldsymbol{u}}
\newcommand{\vv}{\boldsymbol{v}}

\newcommand{\xv}{\boldsymbol{x}}

\newcommand{\jv}{\boldsymbol{j}}
\newcommand{\Rv}{\boldsymbol{R}}
\newcommand{\ev}{\boldsymbol{e}}
\newcommand{\rv}{\boldsymbol{r}}
\newcommand{\drm}{{\rm d}}
\newcommand{\Uv}{{\boldsymbol{U}}}
\newcommand{\Vv}{{\boldsymbol{V}}}

\makeatletter
\def\@cite#1#2{{#1\if@tempswa , #2\fi}}
\def\@biblabel#1{}
\makeatother

\begin{document}

\title{Calculation of Field Line Slippage Rates in Coronal Simulations}

\author[0000-0003-3704-4229]{\underline{Valentin Aslanyan}}
\email{Valentin.Aslanyan@glasgow.ac.uk}
\affiliation{School of Mathematics \& Statistics, University of Glasgow, Glasgow, G12 8QQ, UK}

\author[0000-0003-2297-9312]{David MacTaggart}
\affiliation{School of Mathematics \& Statistics, University of Glasgow, Glasgow, G12 8QQ, UK}
\affiliation{Department of Physics, University of Trento, Povo, 38123, Italy}

\author[0000-0001-8517-4920]{Roger B. Scott}
\affiliation{Montana State University, Bozeman, MT 59715, USA}

\author[0000-0002-0154-8380]{Chloe P. Wilkins}
\affiliation{School of Science, University of Newcastle, University Drive, Callaghan, NSW 2308, Australia}

\author[0000-0002-1089-9270]{David I. Pontin}
\affiliation{School of Science, University of Newcastle, University Drive, Callaghan, NSW 2308, Australia}

\author[0000-0002-2728-4053]{Anthony R. Yeates}
\affiliation{Department of Mathematical Sciences, Durham University, Durham, DH1 3LE, UK}

\author[0000-0002-6442-7818]{Peter F. Wyper}
\affiliation{Department of Mathematical Sciences, Durham University, Durham, DH1 3LE, UK}

\author[0000-0001-6046-2811]{Karen A.~Meyer}
\affiliation{Division of Mathematics, School of Science and Engineering, University of Dundee, Dundee, DD1 4HN, UK}

\author[0000-0003-0176-4312]{Spiro K. Antiochos}
\affiliation{University of Michigan, Department of Climate and Space Sciences and Engineering, Ann Arbor, MI 48109, USA}

\begin{abstract}
We present the Universal Slippage calculator (\texttt{USlip}), an extension of the Universal Fieldline Tracer (\texttt{UFiT}), that computes magnetic field line slippage rates from an input magnetic field defined on a structured grid. The slippage rate provides a local, physically grounded diagnostic of non-ideal evolution, directly quantifying departures from ideal frozen-in motion.  We outline the theoretical formulation and describe an efficient numerical implementation. The method is validated against analytical solutions and applied to simulations using three coronal codes treating varying spatial scales, all of which are standard inputs for \texttt{USlip}. In all the examples considered, the slippage rate not only identifies where magnetic connectivity changes occur, but also accurately reproduces the observed direction of field-line slippage. The consistency of the results across analytical tests and multiple simulation platforms demonstrates that \texttt{USlip} is a robust and versatile tool for diagnosing three-dimensional reconnection in dynamically evolving systems.
\end{abstract}

\section{Introduction}
\label{scn:intro}

The solar corona is a highly structured, magnetically dominated plasma in which the evolution of the magnetic field governs a broad spectrum of dynamic phenomena. A fundamental feature of the coronal magnetic field is its organization into regions of distinct connectivity, commonly described in terms of magnetic topology \citep{Longcope2005}. On sufficiently large scales, and away from strong gradients in connectivity, the evolution of the magnetic field can be approximated by ideal magnetohydrodynamics (MHD), under which magnetic field lines are frozen into the plasma flow.

This approximation breaks down at the boundaries between topological regions. At such locations, sharp gradients or discontinuities in magnetic connectivity are preferential locations for the dynamic development of small scales in the magnetic field -- these enhance the role of non-ideal terms in Ohm’s law, allowing for departures from flux conservation. As a result, magnetic field lines may change their connectivity over time, connecting different plasma elements at different instants. This process, referred to as magnetic reconnection, is central to the dynamics of the solar corona \citep{PontinPriest2022}.

While classical studies of reconnection focused primarily on null-point configurations in two and three dimensions, it is now well established that many coronal reconnection events occur in the absence of magnetic nulls. Such non-null reconnection is inherently three-dimensional (3D) and cannot be fully characterized by null-point theory alone.  Consequently, a range of diagnostic tools has been developed to identify and quantify 3D reconnection processes. These include topological and geometrical measures such as separatrix structures and quasi-separatrix layers \citep[\textit{e.g.}][]{Longcope2005,TitovHornigDemoulin2002,ScottPontinHornig2017}, which indicate regions that exhibit rapid changes in field line connectivity, and which are thus likely sites for reconnection to occur.

However, most existing diagnostics are primarily predictive. That is, they identify locations where reconnection may occur, rather than directly quantifying the local rate at which magnetic field lines depart from ideal evolution. This distinction is particularly important in complex, time-dependent coronal configurations, where reconnection is spatially distributed and dynamically evolving.

To address this issue, the concept of the magnetic field line slippage rate, hereafter the slippage rate \citep{mactaggart2025}, has been introduced as a local, physically grounded measure of non-ideal behavior. The slippage rate quantifies the instantaneous deviation of a magnetic field line from motion that is frozen into the plasma, thereby providing a direct measure of the rate at which field line connectivity changes relative to the bulk plasma velocity (\textit{i.e.} ideal motion). 

As a local diagnostic, the slippage rate complements global topological measures such as the squashing factor $Q$ of QSLs \citep{TitovHornigDemoulin2002} and the related slip-squashing factor \citep{Titov2009SlipSquashing}, which characterize the deformation of field line mappings. Detailed field line mappings, such as the so-called slip-back mapping method, can provide a direct measure of changes in the field line connectivity between two surfaces, and have been used extensively by \citet{Lionello2020} and \citet{Mason2026} for analyzing interchange reconnection in simulations of the evolution of the solar global field. A similar approach has been used by \citet{Dahlin2025}, who used changes in field line length as a measure of reconnection in flare simulations. The issues with these mapping methods is that many field lines need to be computed in order to obtain an accurate measure, which can be numerically costly, and the location of where reconnection occurred between the two surfaces defining the mapping is not determined.  In fact, recent work by \citet{StanishMacTaggart2026} has demonstrated that, whereas $Q$ and the related mapping methods identify regions that are geometrically favorable for reconnection, the slippage rate identifies where reconnection is actively occurring at a given time.

Another important framework for the study of reconnection is general magnetic reconnection \citep[GMR;][]{Schindler1988,Hesse1988}, in which a key diagnostic is the field-line voltage, defined as the integral of the non-ideal electric field along magnetic field lines. While both the field-line voltage and the slippage rate characterize departures from ideal evolution, they may diagnose different aspects of non-ideal behaviour. Appendix \ref{app:gmr} illustrates this distinction through a simple force-free example. Further discussion of the relationship between field-line slippage and GMR may be found in \cite{eyink2015}.

In this work, we extend our previous computational framework, the Universal Fieldline Tracer \citep[\texttt{UFiT};][]{ufitarticle} for magnetic topology, to include the calculation of field line slippage rates. Building on earlier tools for global diagnostics such as connectivity and the squashing factor, we introduce the Universal Slippage calculator (\texttt{USlip}), which computes spatially resolved slippage rate fields from numerical or observational magnetic data.  This enables a unified analysis of magnetic topology and non-ideal dynamics within a single framework.

The purpose of this work is to demonstrate the capabilities of this framework by applying field line slippage diagnostics to a range of representative magnetic configurations. Using the different classes of input supported by \texttt{UFiT}, we show that the approach provides a flexible and robust tool for identifying and quantifying 3D magnetic reconnection in complex datasets.

The layout of this article is as follows: first, we briefly summarize the key results related to the slippage rate, particularly in relation to coronal applications. We then demonstrate the application of \texttt{USlip} to four representative cases, each corresponding to one of the input classes of \texttt{UFiT}.

\section{Theory}
\label{scn:theory}

\subsection{Magnetic field line slippage rate}

The concept of the slippage rate has been introduced and discussed in previous studies \citep{eyink2015,mactaggart2025,StanishMacTaggart2026}. For completeness, however, we briefly summarize the key results required for the present work.

The local (\textit{i.e.} point-wise) evolution of the magnetic field $\Bv$, in relation to the bulk plasma velocity $\uv$, is governed by the induction equation
\begin{equation}\label{eq:induction}
    \frac{\partial\Bv}{\partial t} = \nabla\times(\uv\times\Bv)-\nabla\times\Rv,
\end{equation}
where $\Rv$ represents any non-ideal terms from Ohm's law (\textit{i.e.} $\Ev+\uv\times\Bv=\Rv$, for electric field $\Ev$). At a given time, it may be shown \citep{mactaggart2025} that the instantaneous deviation of a magnetic field line from ideal motion (\textit{i.e.} motion frozen into $\uv$) is given by 
\begin{align}
\Sv = -\frac{(\nabla\times\Rv)_\perp}{|\Bv|}, \label{eqn:sigma}
\end{align}
where  $(\cdot)_\perp$ indicates the projection orthogonal to $\Bv$, defined for any vector $\Vv$ as
\begin{align}
\left(\Vv\right)_\perp \equiv \Vv-\frac{(\Vv\cdot\Bv)\Bv}{|\Bv|^2}.
\end{align}
The slippage rate $\Sv$ has dimensions of $[T]^{-1}$, where $T$ represents time. 

\begin{figure}[h!]
\centering
(a)\includegraphics[scale=0.25]{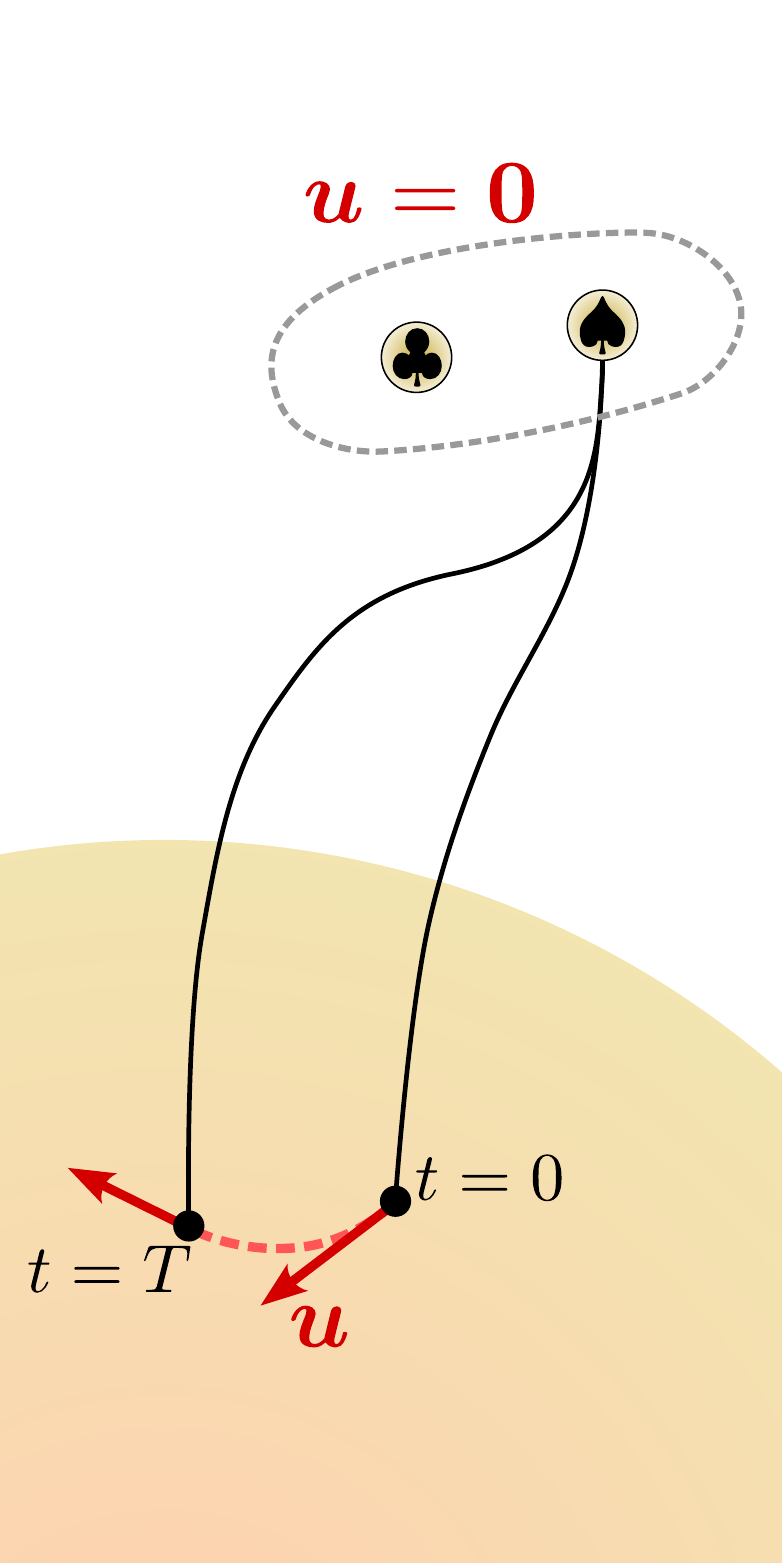} $\;\;$ (b)\includegraphics[scale=0.25]{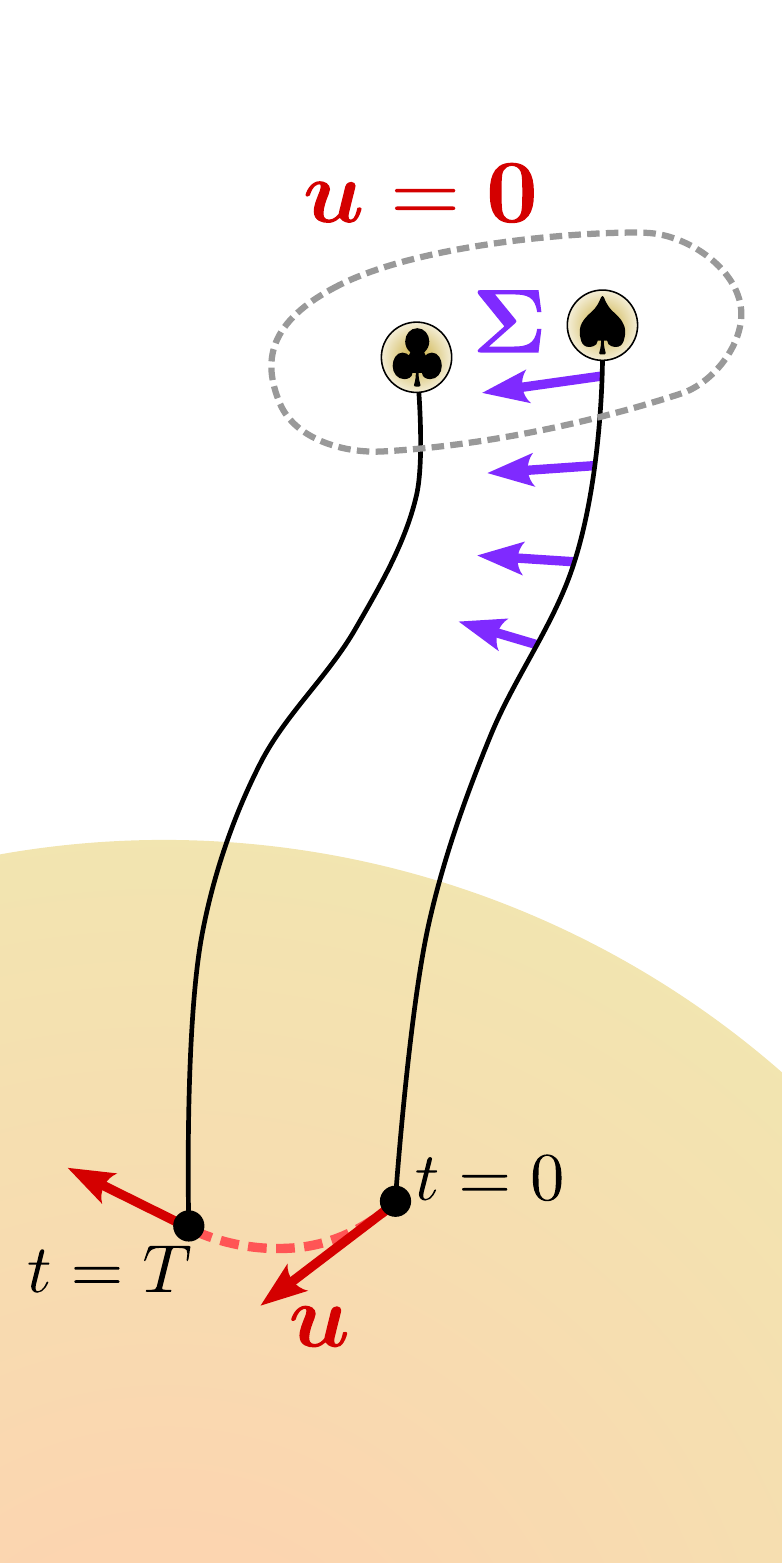}
\caption{Sketch showing two particular blobs of plasma, labelled $(\clubsuit)$ and $(\spadesuit)$, in the solar corona above the photosphere; they are stationary in a region of fluid stagnation, where $\uv=\boldsymbol{0}$. Consider a passive observer particle on the photosphere which lies on a magnetic field line connected initially to $(\spadesuit)$. It moves to a new position at a later time $T$. (a) In the limit of ideal MHD under a smooth velocity field, the field line must remain frozen-in to blob $(\spadesuit)$. Deformation of the field line can be affected only by the velocity field $\uv$. (b) The field line can appear to slip if ideal MHD is violated, so that the observer particle becomes connected to $(\clubsuit)$ at time $T$. The field line slips relative to its ideal motion under the vector field $\Sv$. In this example, $\Sv$ is initially nonzero along the upper portion of the field line, including in the stagnation region. It is the cumulative effect of $\Sv$, {captured by $\vv_{\rm slip}$} in equation \eqref{eqn:sigmaintegral}, that causes the field line to slip from $(\spadesuit)$ to $(\clubsuit)$.}
\label{fig:sketch}
\end{figure}

In applications to the solar corona considered here, we adopt a resistive form for the non‑ideal term 
\begin{equation}\label{R_resistivity}
    \Rv=\eta\jv,
\end{equation} 
where $\eta$ is the resistivity and
\begin{equation}
\jv=\frac{1}{\mu_0}\nabla\times\Bv \label{eqn:currentdensity}
\end{equation} 
is the current density in SI units, with $\mu_0$ being the permeability of free space (other systems of units replace $\mu_0$ with a different constant; see below for examples).  A conceptual illustration of the role of the slippage rate $\Sv$ in 3D reconnection is displayed in Figure \ref{fig:sketch}.

A cumulative measure of the slippage rate along a field line from point $\xv_0$ to point $\xv_1$ may be provided through the slippage ``velocity''
\begin{align}
    \vv_{\mathrm{slip}}(\xv_0,{\xv}_1)=\int_{\xv_0}^{{\xv_1}} \Sv\, {\rm d}\ell \label{eqn:sigmaintegral},
\end{align}
resulting in a quantity with dimensions $[L][T]^{-1}$, where $L$ represents length. The quantity $\vv_{\rm slip}$ is a unique diagnostic of the resultant slippage of a field line relative to ideal motion. This quantity, like $\Sv$ itself, is an instantaneous measure based on the location of the field line at a given instant in time.

An illustration of what $\Sv$ and $\vv_{\rm slip}$ measure is given in Figure \ref{fig:sketch}. In Figure \ref{fig:sketch}(a), the motion satisfies ideal MHD and so, although the field line is dragged along the solar surface, it must remain fixed at $(\spadesuit)$ since $\uv=\mathbf{0}$ at this location. In Figure \ref{fig:sketch}(b), however, the motion is non-ideal and there are non-zero values of $\Sv$ along the field line. The directions of $\Sv$ along the field line result in a non-zero $\vv_{\rm slip}$, encoding that the field line is slipping, relative to ideal motion overall in a particular direction. In this illustrative example, the footpoints of the field line are still dragged at the photosphere but the resultant motion of the field line higher up is to slip from $(\spadesuit)$ to $(\clubsuit)$, and it is slipping relative to ideal motion which, in this case, corresponds to $\uv=\boldsymbol{0}$.

{\subsection{A comparison to flux velocities}

It is useful to compare the slippage rate diagnostics $\Sv$ and $\boldsymbol{v}_{\rm slip}$ with the notion of a flux velocity. A flux velocity $\boldsymbol{w}$ is a velocity field (in this description, constructed from the non-ideal part of Ohm's law) into which magnetic flux is frozen \citep[e.g.][]{HornigSchindler1996,WilmotSmith2005,HornigPriest2003,PriestHornigPontin2003}. Being global quantities, flux velocities are not, in general, guaranteed to exist or be unique.

Even when a flux velocity exists, the information provided by $\Sv$ and $\boldsymbol{v}_{\rm slip}$ can be quite different from that provided by $\boldsymbol{w}$. The examples of \cite{WilmotSmith2005} demonstrate that a non-unique flux velocity may exist while $\Sv=\boldsymbol{0}$ and $\boldsymbol{v}_{\rm slip}=\boldsymbol{0}$.  In these examples, magnetic diffusion changes the field strength but not the geometry of the field lines. Equivalently, the magnetic field direction $\boldsymbol{b} = {\boldsymbol{B}}/{|\boldsymbol{B}|}$ remains fixed, i.e. $ {\partial\boldsymbol{b}}/{\partial t} = \boldsymbol{0}$,  even though the same evolution may be represented through a non-zero flux velocity.  Consequently, and in conjunction with $\uv=\boldsymbol{0}$ for these cases, $(\nabla\times\boldsymbol{R})_\perp =\boldsymbol{0}$ and the slippage rate diagnostics vanish.

This comparison highlights an important distinction. A flux velocity provides a transport description of the magnetic evolution, whereas the slippage diagnostics are determined by $(\nabla\times\boldsymbol{R})_\perp$, the component of the non-ideal evolution associated with changes in the geometry of the magnetic field relative to the bulk plasma velocity. In the examples of \cite{WilmotSmith2005}, this quantity vanishes because the field lines remain unchanged geometrically.

In localized 3D reconnection, a globally defined flux velocity generally does not exist \citep{HornigPriest2003,PriestHornigPontin2003}. Instead, two flux velocities, $\boldsymbol{w}_{\rm in}$ and $\boldsymbol{w}_{\rm out}$, may be constructed by integrating field lines from opposite sides of the diffusion region. Their difference, $\boldsymbol{w}_{\rm out}-\boldsymbol{w}_{\rm in}$, quantifies the mismatch between the two ideal continuations of the magnetic flux through the diffusion region and, hence, the continuous change of field-line connectivity characteristic of 3D reconnection.

Whilst, as indicated above, neither $\Sv$ nor $\boldsymbol{v}_{\rm slip}$ is a flux velocity, $\vv_{\rm slip}$ may identify 3D reconnection in the same way as $\boldsymbol{w}_{\rm out}-\boldsymbol{w}_{\rm in}$. Integrating along field lines that pass through the diffusion region, both $\boldsymbol{w}_{\rm out}-\boldsymbol{w}_{\rm in}$ and $\vv_{\rm slip}$ will typically be non-zero, indicating field line slippage. Calculating these quantities on field lines that do not pass through the diffusion region will lead to them both being zero (the plasma is ideal outside the diffusion region).

Since the slippage diagnostics are derived from local (point-wise) quantities they always exist and are unique. Flux velocities are global and so have more constraints in relation to existence and uniqueness. 

Further, the slippage diagnostics do not require the existence of a localized diffusion region. Examples will be given later in which resistivity is present throughout the plasma and the geometry of the field dictates where $\Sv$ is strongest, thus indicating where field line slippage is most prominent.
}

\subsection{Current density decomposition}

When modelling non‑ideal coronal dynamics using resistive MHD, additional insight into the structure of the slippage rate may be obtained by decomposing the current density according to the local magnetic geometry. Following \cite{mactaggart2025}, we exploit the geometrical properties of the Lorentz force and the field‑aligned twist to write
\begin{equation}\label{eq:cur_dec1}
    \mu_0\jv = \lambda\Bv_{f\perp} + \alpha\Bv,
\end{equation}
 where
\begin{equation}\label{eq:cur_dec_2}
    \Bv_{f\perp} = \Bv\times\ev_F, \quad \lambda = \frac{(\nabla\times\Bv)\cdot\Bv_{f\perp}}{|\Bv|^2}, 
\end{equation}
with unit vector $\ev_F$ in the direction of the Lorentz force $\Fv=\jv\times\Bv$, and
\begin{equation}\label{eq:alpha}
    \quad \alpha = \frac{(\nabla\times\Bv)\cdot\Bv}{|\Bv|^2}. 
\end{equation}
This decomposition separates the current density into two geometrically distinct components: one perpendicular to both the magnetic field and the Lorentz force, and one aligned with the magnetic field itself.  The associated scalar coefficients have clear physical interpretations. The parameter $\lambda$ measures the local rotation of the magnetic field about $\Bv_{f\perp}$, while $\alpha$ corresponds to the familiar field‑aligned twist, representing the local rotation of $\Bv$ about itself.

Substituting Equations (\ref{R_resistivity}) and (\ref{eq:cur_dec1})–(\ref{eq:alpha}) into the definition of the slippage rate in Equation (\ref{eqn:sigma}) yields a series of contributions involving spatial gradients of the resistivity, the Lorentz force, and the field‑aligned twist. The full expression is given in Appendix \ref{appendix:slip_rate}. For applications to large-scale coronal simulations, we focus strictly on uniform $\eta$ in this work.

Of particular importance for coronal applications is the term arising solely from gradients in $\alpha$,
\begin{equation}\label{eq:sigma_alpha}
    \Sv_\alpha=-\frac{\eta}{\mu_0|\Bv|}\nabla\alpha\times\Bv.
\end{equation}
Unlike the remaining contributions, $\Sv_\alpha$ depends exclusively on cross‑field gradients of the field‑aligned twist and is entirely independent of the Lorentz force. This property is especially relevant in the solar corona, which is often close to being force‑free on large scales. In such regions, 3D reconnection is, therefore, directly linked to the spatial structuring of $\alpha$ rather than to strong forces. This connection between the slippage rate and the intrinsic magnetic field geometry has been exploited recently in \cite{StanishMacTaggart2026}, where the magnitude $|\Sv_\alpha|$ is used as a proxy for 3D magnetic reconnection in nonlinear force‑free field extrapolations.

In the remainder of this paper, we focus on the numerical calculation of the full slippage rate $\Sv$ and its force‑free approximation $\Sv_\alpha$, both for constant $\eta$. These are the main quantities determined using \texttt{USlip}. From the decomposition in equation (\ref{eq:cur_dec1}), any differences between these quantities arise from Lorentz force‑related effects.

\section{Numerical calculation of the magnetic slippage rate}
\label{sec:uslip}

The calculation within \texttt{USLip} of the magnetic field line slippage rate $\Sv$ and its force‑free approximation $\Sv_\alpha$ forms part of the broader \texttt{UFiT} framework \citep{ufitarticle}. \texttt{UFiT} is designed as a post‑processing tool and is able to ingest magnetic field data that are either supplied directly by the user, computed internally using the Potential Field Source Surface (PFSS) model \citep{Schatten1969,virtanen2020,knizhnik24} from an input magnetogram or produced by a range of widely used numerical magnetohydrodynamic and magnetofrictional solvers, including the \texttt{ARMS} \citep{devore1991}, \texttt{Lare3D} \citep{lare3d} and \texttt{DUMFRIC} \citep{dumfric} codes. The package is entirely self‑contained and does not require any of these external codes to be installed locally, facilitating straightforward offline analysis.

The \texttt{USlip} package calculates the following scaled diagnostic quantities
\begin{align}
    \jv^* &= \nabla \times \Bv, \\
    \Sv^* &= -\frac{\left[\nabla \times \left(\nabla \times \Bv \right)\right]_\perp}{|\Bv|}, \\
    \Sv_\alpha^* &= -\frac{1}{|\Bv|} \nabla \left[\frac{\left(\nabla \times \Bv \right)\cdot \Bv}{|\Bv|^2}\right]\times \Bv,
\end{align}
which are returned on the same spatial grid as the input. The relative magnitudes of $\Sv^*$ and $\Sv_\alpha^*$ indicate regions of rapid slippage rate, while their directions indicate the directions of field line slippage. It supports Cartesian, spherical, and cylindrical coordinates (see Appendix \ref{scn:ufitcyl} for the method of calculating the squashing factor in the latter).

The physical units of the magnetic field are not required for operations based solely on magnetic topology, such as field line tracing or squashing factor calculations (in \texttt{UFiT}) and are, therefore, ignored in those contexts. For the computation of absolute slippage rates, however, the system of units is important. The scaled outputs of \texttt{USlip} are related to their dimensional counterparts by 
\begin{align}
    \jv &= \frac{1}{\mu_0 L}\jv^*, \label{eqn:jstar} \\ 
    \Sv &=\frac{\eta}{\mu_0 L^2}\Sv^*, \label{eqn:svstar} \\
    \Sv_\alpha&=\frac{\eta}{\mu_0 L^2}\Sv_\alpha^*, \label{eqn:svalphastar} 
\end{align}
where $L$ is the length scale of the supplied grid (\textit{e.g.} spherical grids may be implicitly in units of $R_\odot$), $\mu_0$ is the constant in the definition of the current density and $\eta$ the plasma resistivity.

One factor of $L$ appears for every derivative taken, as \texttt{USlip} uses the intrinsic grid units. The scaled current density should be multiplied by a uniform factor of $1/\mu_0$, which depends on the units of the input $\Bv$, to obtain the true current density. The two slippage rates do not require the units of $\Bv$ to be known, as both are normalized by the field strength; instead, multiplication by a factor of $\eta/\mu_0$ (also independent of electromagnetic units) is used to obtain a true slippage rate. This requires a value of the resistivity $\eta$ to be specified. In realistic plasmas, this may depend on properties such as density $\rho$ and temperature $T$, while in simulations numerical resistivity may also depend on grid spacing $\Delta x$. \texttt{USlip} returns its outputs directly on the input grid, and therefore the fully dimensional values of $\Sv$ and $\Sv_\alpha$ are obtained through multiplication by $\eta$ defined on that same input grid; it may be known \textit{a priori} (for example, as a simulation parameter), or computed using local simulation parameters from first principles.

To compute the slippage rate diagnostics, \texttt{USlip} first obtains the full magnetic field $\Bv$ on a structured grid. In many numerical codes, the components of $\Bv$ are stored on staggered grids (for example, face‑centred or cell‑centred discretizations) in order to improve numerical stability or accuracy during time integration. Since the definitions of the slippage rates involve vector norms and derivatives that assume co‑located field components, \texttt{USlip} first regularizes the magnetic field onto a common grid, in line with how this is done by the codes themselves. This step is essential for evaluating the normalizations appearing in equations (\ref{eqn:sigma}) and (\ref{eq:sigma_alpha}), and ensures consistency across different simulation outputs.

Once the magnetic field has been regularized, \texttt{USlip} performs two sequential passes over the grid. In the first pass, spatial derivatives of $\Bv$ are calculated in order to evaluate $\nabla\times\Bv$ and the field‑aligned twist $\alpha$. In the second pass, these quantities are used to compute the full slippage rate $\Sv$ and its force‑free approximation $\Sv_\alpha$, with each derivative evaluated only once per pass. Where available, shared‑memory parallelism is exploited via OpenMP to accelerate the calculation.

Spatial derivatives are evaluated using finite differences in the coordinate system appropriate to the input data: Cartesian, $(x,y,z)$; spherical $(r,\theta,\phi)$; cylindrical $(r,\phi,z)$. In \texttt{ARMS} simulations, the magnetic field is sometimes defined on a spherical exponential grid, with $\ln(r)$ stored as the radial coordinate. In such cases, \texttt{USlip} automatically converts the grid to physical radius $r$ prior to computing derivatives.

For a grid point indexed by coordinates $(i,j,k)$ in the chosen coordinate system, e.g. $(x_i,y_j,z_k)$ for Cartesian coordinates, \texttt{USlip} identifies all six adjacent points $(i\pm 1,j\pm 1,k\pm 1)$ whenever possible. Away from non‑periodic boundaries, first derivatives are then evaluated using second-order central‑difference formulae. At non‑periodic boundaries, appropriate first-order forward or backward differences are used instead. Further details are given in Appendix \ref{scn:derivatives}. Note that this approach avoids the need to interpolate directly, as derivatives of the magnetic field are computed at the same grid points where the field is defined precisely. Interpolation may be used for post-processing tasks, such as visualization or calculation of slippage along field lines as shown in Section \ref{scn:simulations}.

\begin{figure}[h!]
\centering
\includegraphics[scale=0.3]{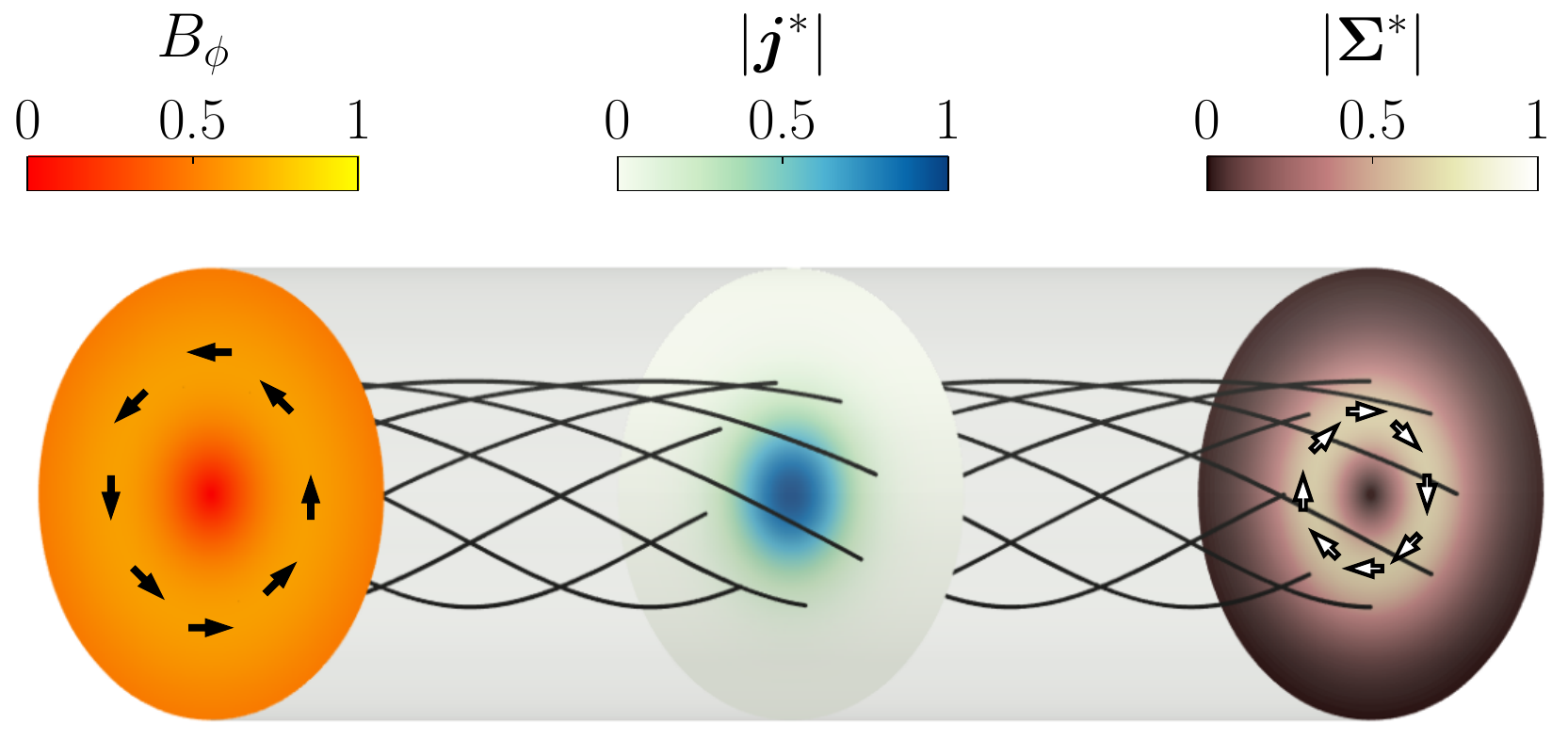}
\caption{Magnitudes of the non-axial component of the magnetic field, the axial current density, and the resulting field line slippage in a magnetized Z-pinch with $B_0=l=1$, $C=0.5$. Solid black arrows indicate the direction of the field, while white-filled arrows indicate the direction of the slippage. These quantities are constant along the $z$-axis (left to right in this figure). The field lines slip in the opposite direction to the field. The current and slippage, as well as the trajectories of the field lines are computed by the \texttt{UFiT} package (including \texttt{USlip}).}
\label{fig:zpinch}
\end{figure}

\subsection{Analytical test}
\label{scn:zpinch}

To validate the numerical implementation of the slippage rate diagnostics, we consider a simple analytical test problem that admits closed‑form solutions in both Cartesian and cylindrical geometries. Specifically, we examine a magnetized Z-pinch (sometimes referred to as a ``screw pinch'') configuration in which a strong, spatially localized current flows parallel to a uniform background magnetic field.

In cylindrical coordinates $(r,\phi,z)$, the background magnetic field is taken to be $B_0\ev_z$, together with an imposed current density
\begin{align}
\jv = \frac{j_0}{\mu_0}e^{-r^2/l^2}\ev_z, \label{eqn:zpinchj}
\end{align}
where $j_0$ and $l$ are constants. The total magnetic field can then be written as $\Bv = B_\phi(r)\ev_\phi + B_0\ev_z$, where the $\phi$-component must satisfy $\mu_0\jv =\nabla\times\Bv$. In cylindrical coordinates, this yields
\begin{align}
\frac{\drm}{\drm r}(rB_\phi(r)) = j_0re^{-r^2/l^2},
\end{align}
which can be integrated to give
\begin{align}
B_\phi(r) &=C\frac{1-e^{-r^2/l^2}}{r}, \label{eqn:zpinchbcyl}
\end{align}
where $C=j_0l^2/2$. Note that a Taylor expansion shows that this expression is not singular at $r=0$.

The corresponding slippage rate may be written in closed form as
\begin{equation}
    \Sv^* = -\frac{WB_0^2}{(B_\phi^2+B_0^2)^{3/2}}\ev_\phi + \frac{W B_\phi B_0}{(B_\phi^2+B_0^2)^{3/2}}\ev_z,
\end{equation}
where $W=2 r j_0\exp(-r^2/l^2)/l^2$. These expressions can be straightforwardly recast in Cartesian coordinates by noting that $r=\sqrt{x^2+y^2}$ and $\ev_\phi = -\ev_x y/r+\ev_y x/r$.

This configuration requires only the magnetic field to be defined on a grid and, therefore, provides a clean test of the numerical procedure; in this case, the grid has equal spacing in each dimension. Using the analytically constructed field as input, \texttt{USlip} accurately reproduces the scaled quantities $\jv^*$, $\Sv^*$, and $\Sv_\alpha^*$ in both Cartesian and cylindrical coordinates. Representative results are shown in Figure \ref{fig:zpinch} for the set of parameters $B_0=l=1$, $C=0.5$. In this example, a single \texttt{Python} script is sufficient to construct the magnetic field, invoke \texttt{USlip}, and visualize the output, without external input.

\begin{figure}[h!]
\centering
\includegraphics[scale=0.46]{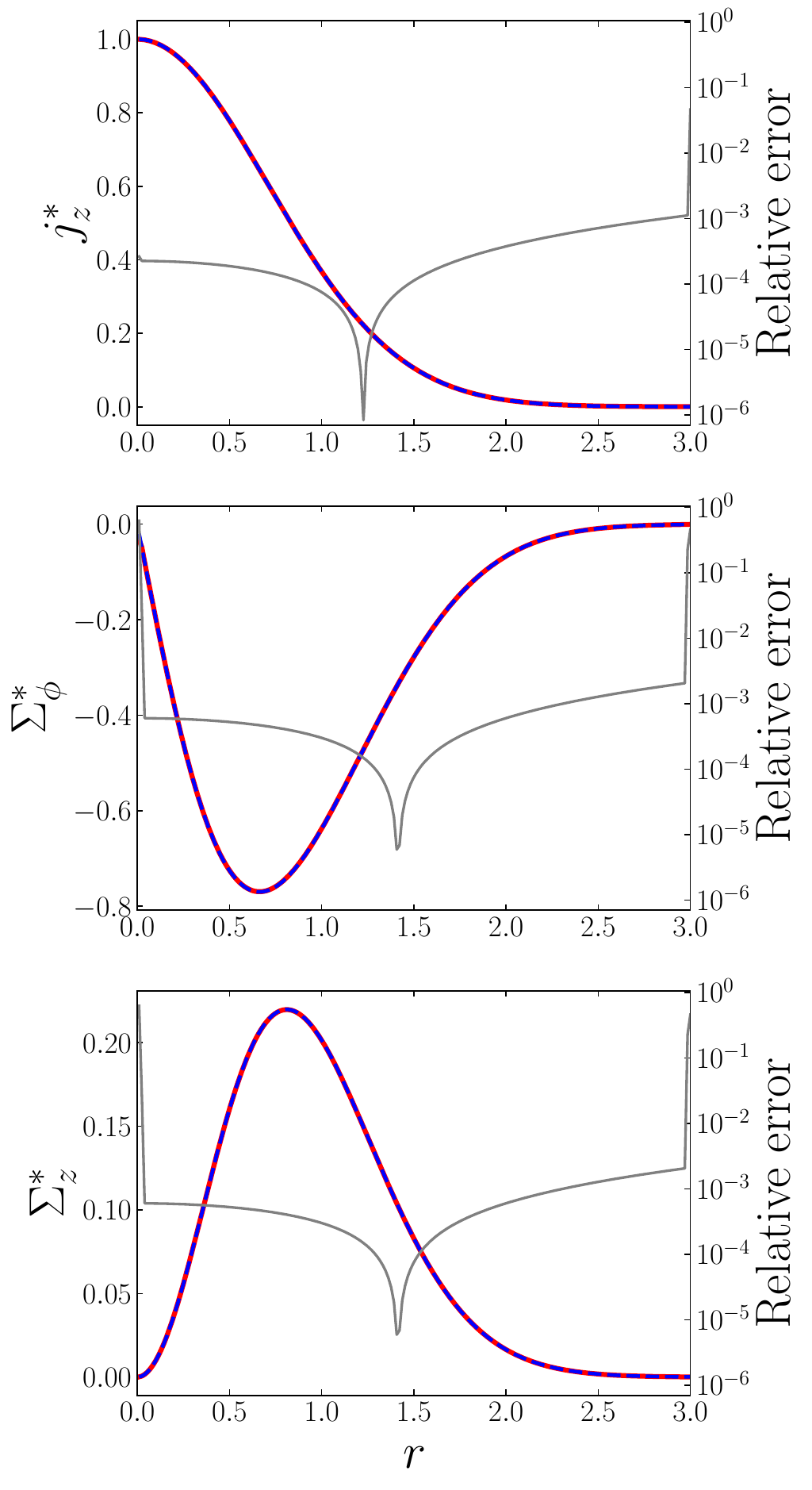}
\caption{Comparison of \texttt{USlip} results (solid, blue) with the corresponding analytic formulae (dashed, red) for the magnetized Z-pinch; the scale is on the left. The light gray line shows the relative error, \textit{e.g.} $|j_z^*-j_{z,\mathrm{analytic}}^*|/|j_{z,\mathrm{analytic}}^*|$; the error scale is on the right. The error is minimized in the middle of the domain, but rises sharply at both end points of the grid.}
\label{fig:accuracy}
\end{figure}

The resulting field lines are helical and are found to slip in the direction opposite to their pitch (\textit{i.e.} $B_\phi$ and $\Sigma_\phi$ have opposite signs), in precise agreement with the analytical solution. A quantitative assessment of the accuracy of the numerical calculation, and those of the other quantities produced by \texttt{USlip} for this example, is shown in Figure \ref{fig:accuracy}. For most of the domain, the relative error $|j_z^*-j_{z,\mathrm{analytic}}^*|/|j_{z,\mathrm{analytic}}^*| \leq 10^{-3}$, with increased error at the grid edges, where a lower finite difference formula is used, while the error is minimized in the middle of the domain. Furthermore, $\Sigma^*_{\phi,\mathrm{analytic}}\rightarrow 0$ and $\Sigma^*_{z,\mathrm{analytic}}\rightarrow 0$ at the inner grid edge. Increased spatial resolution reduces the relative error.

\section{Examples of magnetic slippage in simulations}
\label{scn:simulations}

In this section, we illustrate the application of \texttt{USLip} to the analysis of magnetic reconnection in outputs from simulations using three codes currently supported by \texttt{UFiT} (in addition to fields passed in as a \texttt{UFiT}-specific binary file, or directly as a \texttt{numpy} array in \texttt{Python}): \texttt{ARMS}, \texttt{DUMFRIC} and \texttt{Lare3D}.

\subsection{Field lines under supergranular motion (\texttt{ARMS})}
\label{scn:arms}

\begin{figure}[h!]
\centering
\includegraphics[scale=0.2]{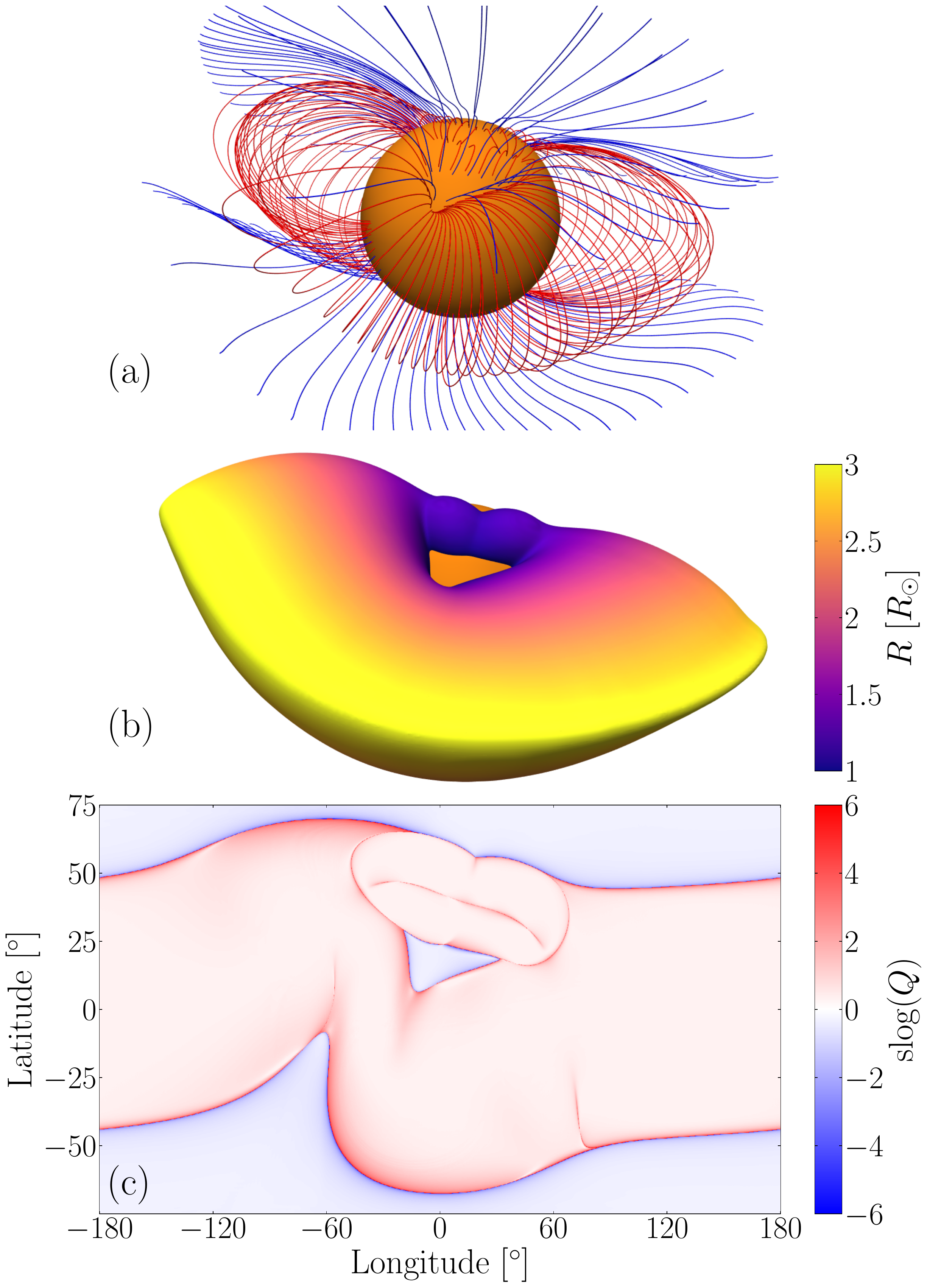}
\caption{At the start of the fully spherical simulations with the \texttt{ARMS} code described here: (a) sample of open (blue) and closed (red) magnetic field lines around a mid-latitude coronal hole; (b) corresponding Last Closed Flux Surface, which separates open and closed field lines; (c) map of the squashing factor $\mathrm{slog} (Q)\equiv \pm \log |Q|$ at $r=R_\odot$, where $(+)$ denotes closed and $(-)$ denotes open field lines.}
\label{fig:armssetup}
\end{figure}

We have previously reported simulations of the dynamics of a coronal hole, the boundary of which is driven by supergranular motion, in \cite{aslanyan2021}. The initial magnetic field is based on one considered by \cite{titov2011}, in which a region of open magnetic field lines is bounded from the south by a streamer belt, and from the north by a pseudostreamer arising from parasitic flux. We show a selection of open and closed magnetic field lines near the boundary of the mid-latitude coronal hole under consideration here, as well as the two ordinary polar coronal holes at the poles in Figure \ref{fig:armssetup}. The 3D surface which separates these two classes of field lines, which we call the Last Closed Flux Surface, is also shown.

One tool used to analyze the magnetic geometry is the squashing factor $Q$, which is shown for the inital time $t=0$ in Figure \ref{fig:armssetup}(c). This quantity is a measure of the non-local mapping of field lines \citep{TitovHornigDemoulin2002,ScottPontinHornig2017} with larger values corresponding to larger deformations. We use the sign of $Q$ to denote the ultimate connectivity, where $Q>0$ denotes closed field lines (both ends terminate on the photosphere) and $Q<0$ denotes all others, including open (precisely one end at the photosphere) and fully disconnected field lines.

\begin{figure*}[t!]
\centering
\includegraphics[scale=0.25]{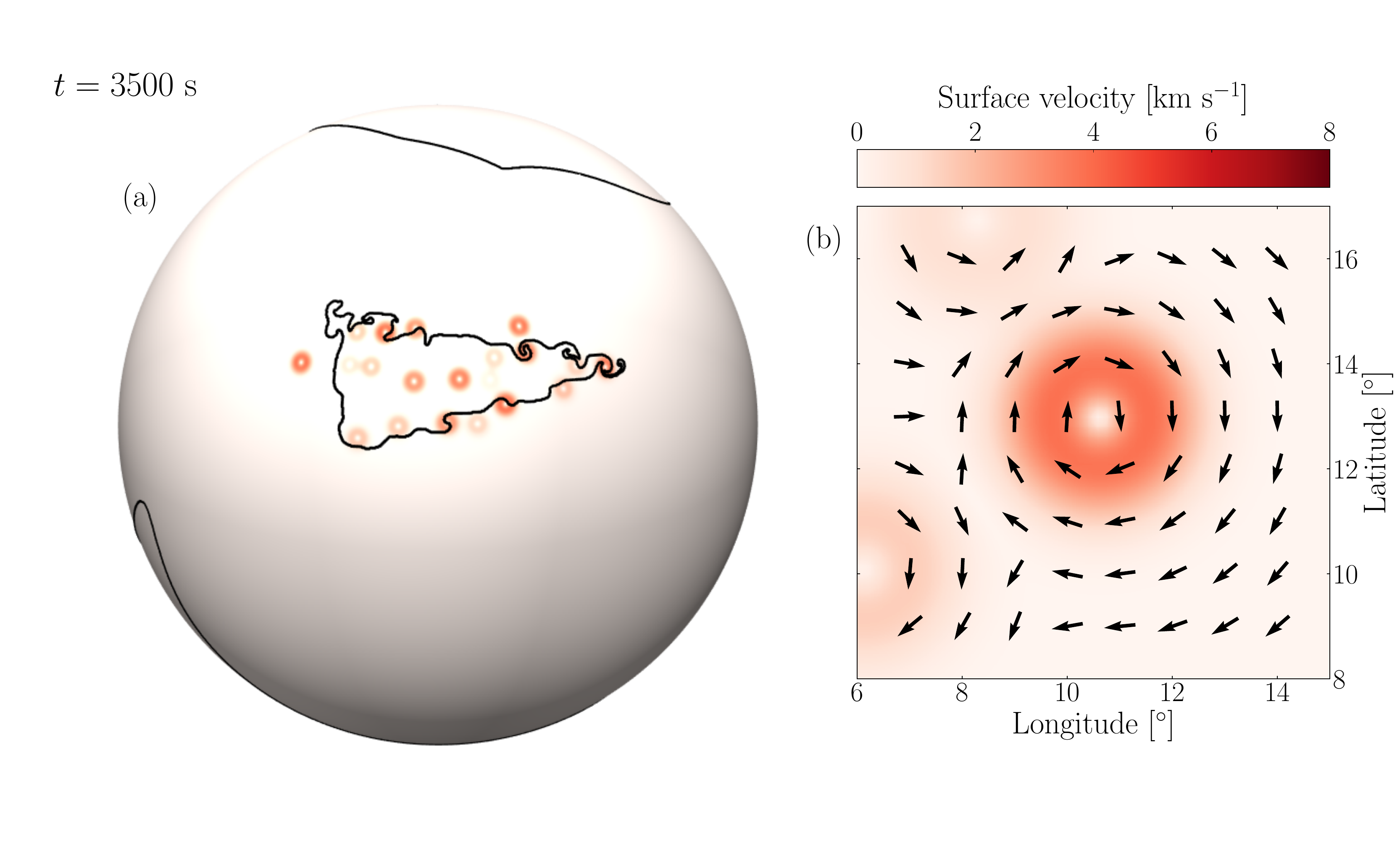}
\caption{(a) Magnitude of the surface velocity boundary condition over the surface of the sun at $r=R_\odot$, approximately mid-way through the simulation at $t=3500$~s, indicated by the red color. The black line indicates the instantaneous coronal hole boundary (separating the footpoints of open and closed field lines at that time). (b) Detail of one supergranular vortex, including the flow direction indicated by the arrows. The animated version of this figure shows these quantities in time from $t=0$ to $t=6000$~s at times indicated. Note that at later times, the coronal hole boundary continues to move; this would be forbidden by ideal MHD and is due to slippage and reconnection discussed in the text.}
\label{fig:surfacev}
\end{figure*}

The \texttt{ARMS} code is used for this study. The corona is assumed to be constantly isothermal at $T=2$ MK. The starting magnetic field of Figure \ref{fig:armssetup} is imposed initially and allowed to relax for $5000$ s of simulation time. The non-ideal effects result purely from numerical resistivity, which is inherent in any numerical solver. MHD variables in the \texttt{ARMS} code are stored in blocks of $8\times 8 \times 8$ points, tiled into an entire grid. The resolution in selected regions of space can be refined using the \texttt{Paramesh} toolkit \citep{MACNEICE2000330}; this is done by taking an existing block and subdividing it into two along each dimension (therefore, evenly breaking up one block into eight). The refinement can be performed recursively, therefore doubling the resolution locally for every level. We have manually chosen to refine the grid in the region of the dynamically important pseudostreamer by a factor of $4$ (hence a $16$-fold linear resolution increase) at the start of the simulation, and thereafter disabled dynamic refinement to simplify the following analysis.

We consider the setup and initial relaxation period to end at $t=0$ and thereafter we drive the system using surface flows resembling supergranules, albeit larger and more rapidly moving than solar values due to computational limitations. The surface flows are divergence-free, taking the form of
\begin{align}
v_\theta&=v_i \mathbb{G}(\theta-\theta_{c,i})\mathbb{G}'(\phi-\phi_{c,i})\frac{1}{\sin\theta}f_i(t), \label{eqn:vth} \\
v_\phi&=-v_i \mathbb{G}(\phi-\phi_{c,i})\mathbb{G}'(\theta-\theta_{c,i})f_i(t), \label{eqn:vph}
\end{align}
where the Gaussian function $\mathbb{G}(x)\equiv \exp(-cx^2)$ and its derivative $\mathbb{G}'(x)\equiv d\mathbb{G}/dx$. For the $i^{\rm th}$ flow, $v_i$ is a constant (positive or negative), $\theta_{c,i}$ and $\phi_{c,i}$ are coordinates of the flow center. Each flow has a time envelope
\begin{equation}
f_i(t)=\frac{1}{2}\left[1-\cos\left(\frac{2\pi (t-t_i)}{T}\right)\right]\quad\;\; t_i<t<t_i+T
\end{equation}
with individual start time $t_i$ and a period $T=2000$~s for all flows. Thus, each of the surface flows is wholly either clockwise or anti-clockwise, but equal numbers of both types of flow are selected so that total helicity injected is approximately zero. The magnitude of the time dependent surface velocity is shown in Figure \ref{fig:surfacev}, as well as a detail of a single supergranule.

The surface motion transports flux and twists up magnetic field lines. Its effect, therefore, qualitatively resembles the situation in Figure \ref{fig:sketch}, whereby footpoints at the photosphere are moved rapidly, while the field lines they seed connect to almost stationary plasma far above. A line showing the instantaneous open/closed boundary on the solar surface is also shown in Figure \ref{fig:surfacev}. It is to be expected that the surface motion deforms this open/closed boundary as, for example, some open field lines may be transported south, while corresponding closed field lines are transported north. In the limit of ideal MHD, however, these field line motions should only occur due to driving at the photosphere and conversely should persist unaltered when the velocity there returns to zero. Instead, as we reported previously, interchange reconnection occurs between the two classes of field lines, which causes the open/closed boundary to smooth out and continue deforming after the driving is switched off, as seen in the animated version of Figure \ref{fig:surfacev}. We now have local diagnostics for the rate of the non-ideal motion of the field lines in the form of $\Sv$ and $\Sv_{\alpha}$.

\begin{figure*}[t!]
\centering
\includegraphics[scale=0.22]{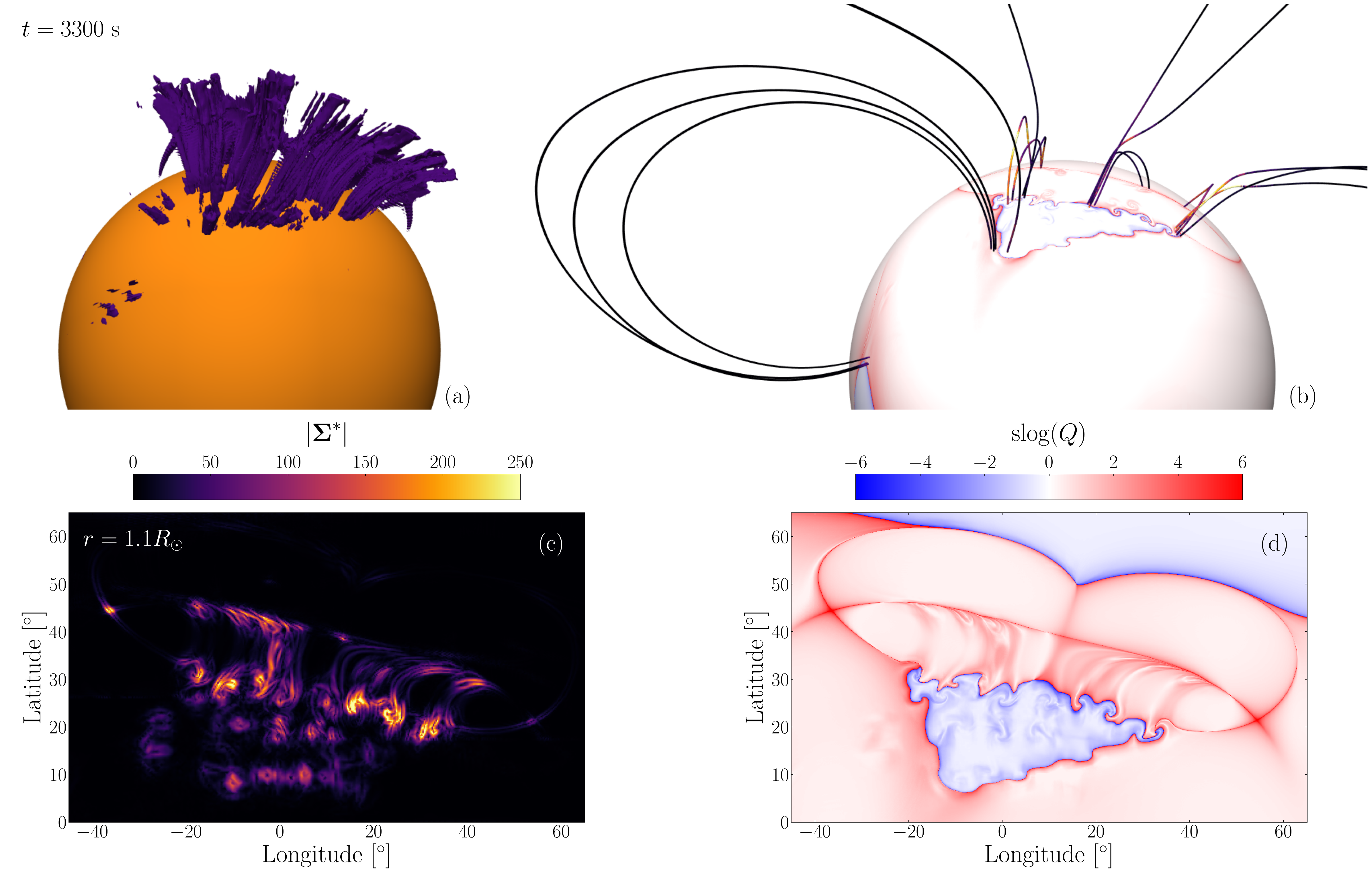}
\caption{(a) Isosurface of the slippage rate at $|\Sv^*|=50$ above the sun sphere. (b) A selection of field lines, colored by the slippage rate, above the sun sphere, colored by the squashing factor. (c) Slippage rate and (d) signed logarithm of the squashing factor in a slice at a constant radius $r=1.1 R_\odot$. Note that the slippage color scale is consistent across panels (a), (b) and (c). This is the situation at $t=3500$~s, while the animated version of this figure shows these quantities in time from $t=0$ to $t=6000$~s at times indicated. The footpoints of the field lines in (b) are advected in time by the surface motion shown in Figure \ref{fig:surfacev}.}
\label{fig:armssigma}
\end{figure*}

Given that \texttt{ARMS} relies purely on numerical resistivity for non-ideal effects, we assume that the resistivity $\eta$ is linearly proportional to the grid spacing. This is in turn determined by the local grid refinement. We, therefore, assume a resisitvity
\begin{align}
\eta =  2^{-\mathcal{R}} \eta_0,
\end{align}
where $\mathcal{R}$ is the local refinement level and $\eta_0$ is some constant to be determined. This relative variation in refinement has been compensated for in the following section. Note also that length measurements in \texttt{ARMS} are scaled to solar radii, so that $L=R_\odot \approx 6.96\times 10^{10}$~cm in equations (\ref{eqn:jstar}) to (\ref{eqn:svalphastar}). \texttt{ARMS} uses the CGS system of units where $\jv = \left(\nabla \times \Bv \right)/4\pi$, in units of abA~cm$^{-2}$; $\eta_0$ is defined in units of ab$\Omega$~cm. The scaled quantities output by \texttt{USlip} are therefore related to their dimensional counterparts by
\begin{align}
    \jv &= \frac{1}{4\pi R_\odot} \jv^*,\label{eqn:armsscale1} \\
    \Sv &=\frac{\eta_0}{4\pi R_\odot^2}\Sv^*,\label{eqn:armsscale2} \\
    \Sv_\alpha &=\frac{\eta_0}{4\pi R_\odot^2 } \Sv_\alpha^*. \label{eqn:armsscale3}
\end{align}

The supergranular nature of the flows causes the twisting up of field lines in roughly cylindrical bundles. An isosurface of constant $|\Sv^*|=50$ part way through the simulation, shown in Figure \ref{fig:armssigma}(a), comprises of tubes of high slippage rate extending upwards from each supergranule. Figure \ref{fig:armssigma}(b) depicts field lines colored by $|\Sv^*|$ along their length, showing exactly where a given field line slips most. For example, in some cases a short pseudostreamer arc may slip near each footpoint, but not at its apex. The animated version of this figure shows the dynamic situation at every $100$~s. The footpoints of certain field lines are initialized at $t=0$ and thereafter advected along the surface according to the known imposed velocity field discussed above; field lines are then traced upwards at subsequent times. The field lines appear to deform from those places along their length where the slippage rate is higher (brighter segments). Selected field lines at either end of the pseudostreamer are seen to undergo interchange reconnection due to their slippage, as depicted.

A slice of the magnitude $|\Sv ^*|$ at a constant radius is shown in Figure \ref{fig:armssigma}(c) at $t=3300$~s, and at every $100$~s in the animated version. In its southern portion, where the field lines are close to radial, there are regions of high slippage rate which are round and hollow, very similar to the underlying supergranular drivers; the slippage therefore forms radial tubes here. In the northern portion, above the pseudostreamer where the field lines curve sharply, these tubes are likewise curved; they are bisected by a slice at constant radius at an oblique angle, resulting in long filamentary structures. At later times when the surface driving is switched off, the slippage rate begins to fall as the system relaxes to a new equilibrium.

\begin{figure}[h!]
\centering
\includegraphics[scale=0.25]{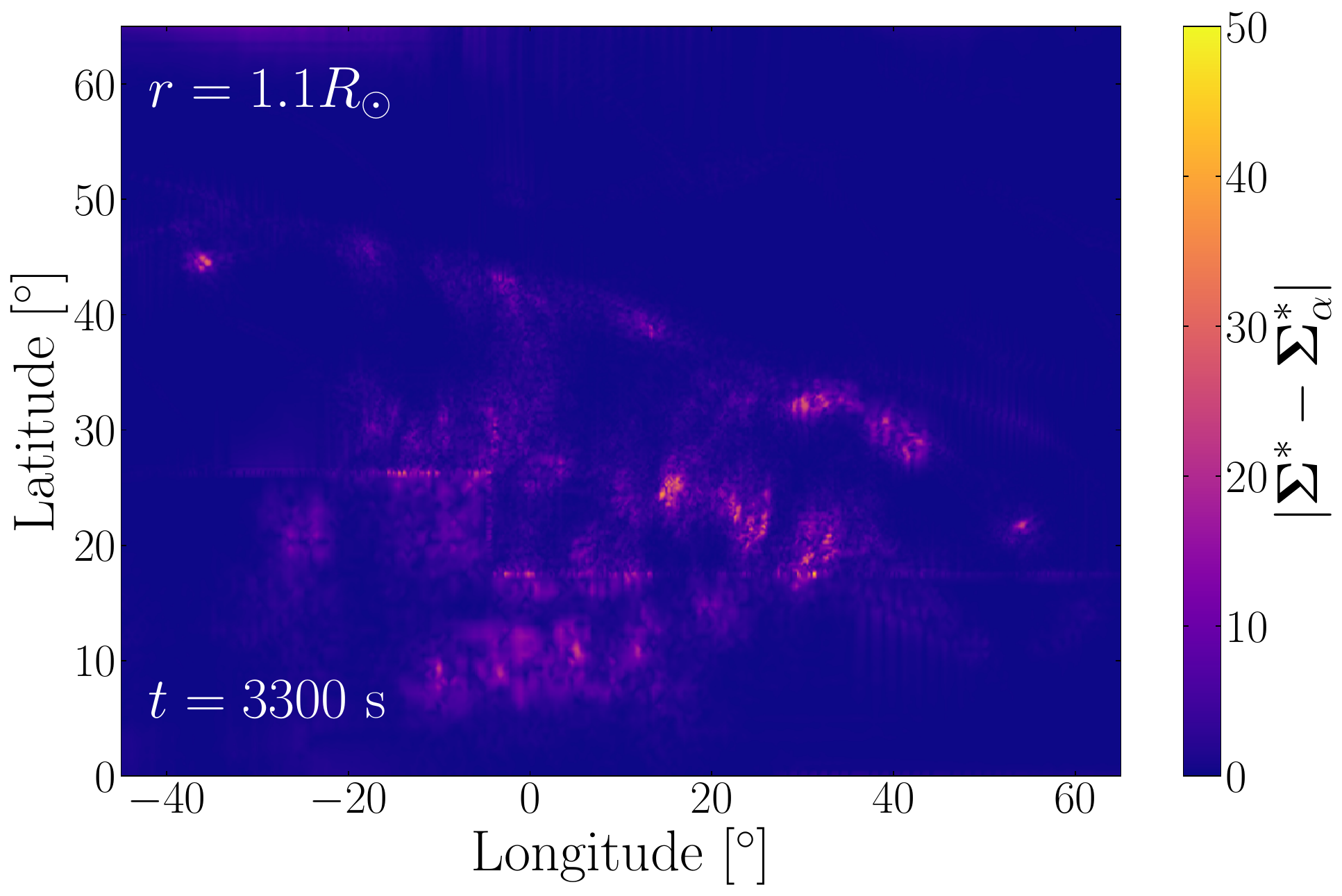}
\caption{During the \texttt{ARMS} simulation, magnitude of the difference between the vector fields of full slippage, and slippage due to field aligned currents at $t=3300$~s. The animated version of this figure shows these quantities in time from $t=0$ to $t=6000$~s at times indicated.}
\label{fig:armssigcomp}
\end{figure}

Figure \ref{fig:armssigcomp} displays $|\Sv^*-\Sv^*_\alpha|$ at 1.1 $R_\odot$ and $t=3300$ s, analogous to the map in Figure \ref{fig:armssigma}(c). This difference map shows that cross-field gradients in $\alpha$ provide the dominant contribution to the slippage rate magnitude. Even in locations where the difference between the full slippage rate and its force-free approximation is most pronounced, it remains less than approximately one fifth of the maximum value of $|\Sv^*|$. Thus, the contribution omitted from $\Sv_\alpha^*$ generally acts as a correction to the full slippage rate rather than contributing at the same order of magnitude. This behaviour is consistent with the coronal magnetic field being close to force-free on large scales. Note that some artifacts are present due to a jump in grid refinement.

The supergranular driving creates a combination of fields and currents qualitatively similar to those in Section \ref{scn:zpinch} and Figure \ref{fig:zpinch}. Initially straight field lines are driven helical, a centrally-peaked current is established, and the resulting field line slippage rate is, therefore, directed around this current.

\begin{figure*}[p!]
\centering
\includegraphics[scale=0.28]{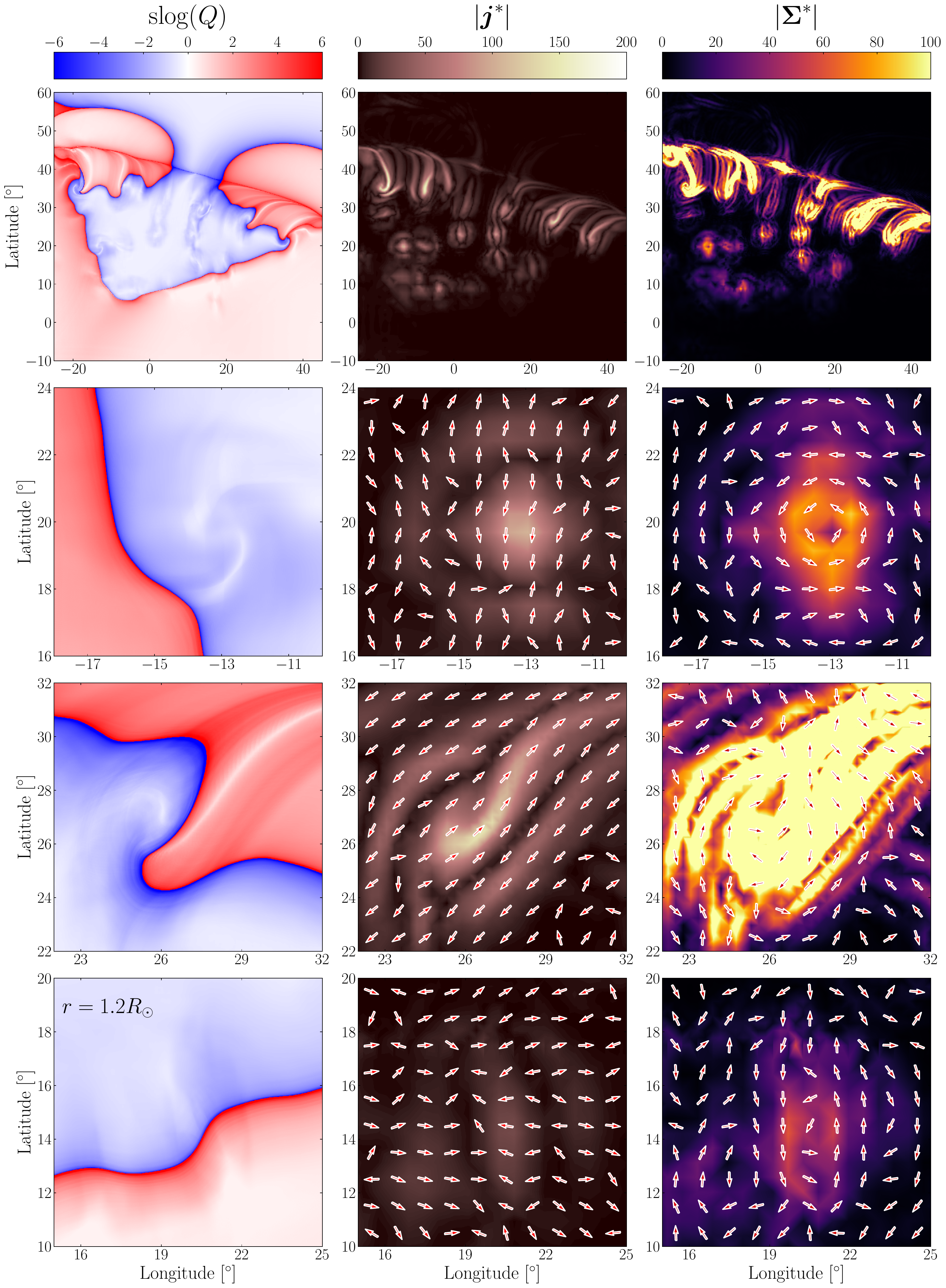}
\caption{Slices of squashing factor (left column), scaled current density (center column) and scaled slippage rate (right column) at constant radius $r=1.2R_\odot$ at time $t=1500$~s, after driving by the initial supergranules has begun. The arrows indicate the direction of the components of the appropriate vector field within these slices.}
\label{fig:arms3diag}
\end{figure*}

We compare the slippage rate to two other key diagnostics for magnetic reconnection and non-ideal behavior -- the squashing factor and the current density -- in Figure \ref{fig:arms3diag}. The top row of panels gives an overview above the entire coronal hole under consideration, while the lower three rows show regions of interest. At a time $t=1500$~s, the supergranular driving has begun to advect field lines, alter the field topology and establish currents.

The second row from the top shows the region around $r=1.2R_\odot,\;\theta=20^\circ,\;\phi=-14^\circ$, above the middle left of the coronal hole. The field lines are approximately radial there. The squashing factor shows that the coronal hole boundary has been deformed. Moreover, the surface flows have created an ``S''-shaped disturbance in the squashing factor due to the twist imparted upon the field lines. In the panels depicting the current density and slippage rate, the arrows indicate their respective direction within a surface of constant radius (\textit{i.e.} excluding the radial component; the magnitude still includes it, however). The direction of the current flow is not meaningful here, but the slippage rate shows a clear anticlockwise circular motion, as driven by the surface flow.

The third row from the top, centered around $r=1.2R_\odot,\;\theta=26^\circ,\;\phi=26^\circ$, is located at the pseudostreamer, where the field lines are strongly inclined. The slice at constant radius therefore intersects the field lines at an oblique angle. The projection, therefore, indicates the direction of the current more clearly; a strong current tube is flowing diagonally upwards and to the right, flanked by two return flows opposite. Resolved in this way, the field lines are therefore slipping over the top of this structure as to be expected from Equation (\ref{eqn:sigma}).

The lowest row of panels in Figure \ref{fig:arms3diag} is in a region above two neighboring vortices, so there is no single clear circular pattern in any of the diagnostics. Nonetheless, the slippage rate shows a rough oval ``O''-shape, with counterclockwise motion around it.

\begin{figure}[t!]
\centering
\includegraphics[scale=0.28]{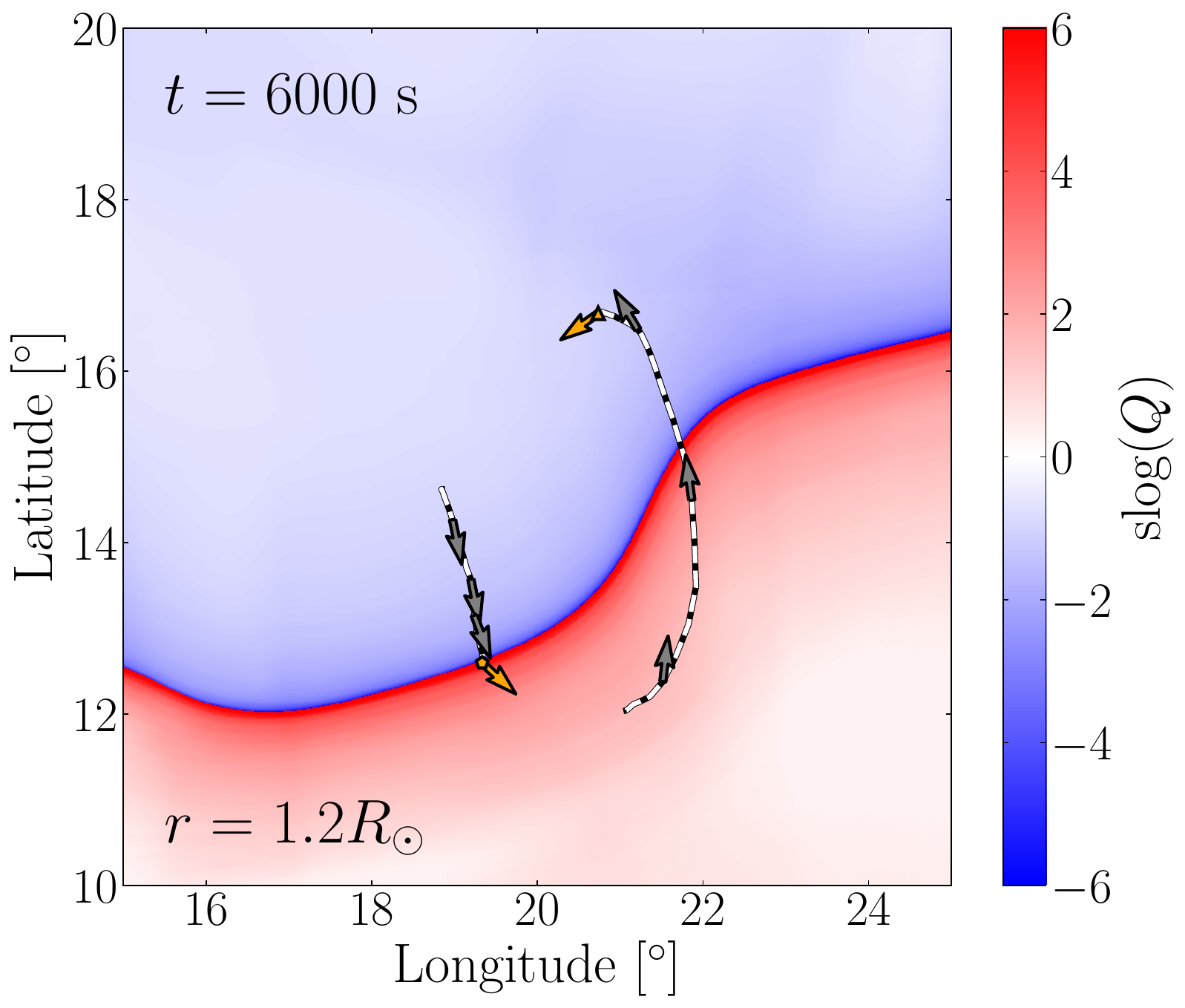}
\caption{Slice of the squashing factor $Q$ at constant radius $r=1.2R_\odot$ at the end of the simulation, $t=6000$~s. The footpoints of two field lines have been ideally advected on the photosphere ($r=R_\odot$), and the path where those field lines intersect this radial slice is given by the dashed line. The direction of the slippage vector field $\Sv$ is indicated by the orange arrow at the end, and by grey arrows part way through the simulation. The animated version of this figure displays $Q$ at times from $t=0$ to $t=6000$~s as indicated; the path of the field lines is shown from the start up to that point in time, and the instantaneous $\Sv$ direction is denoted by the orange arrows.}
\label{fig:armsarrows}
\end{figure}

One key advantage of the slippage rate over the other two diagnostics is that, in addition to showing where non-ideal effects for the field line motion are strongest, it gives a clear indication of the direction in which the slip actually happens. Let us examine the counterclockwise slippage around the ``O''-shape mentioned above in more detail. Consider a footpoint on the photosphere, which moves ideally under the imposed surface velocity shown in Figure \ref{fig:surfacev}. A magnetic field line can be traced into the corona from wherever this footpoint is located. If the field line is long enough, it will intersect a sphere at $r=1.2R_\odot$ at a given time, and therefore form a trajectory over the course of the simulation. From the point of view of the footpoint, the resulting trajectory is quite arbitrary, and a function of the space- and time-dependent magnetic field. A pair of such trajectories are shown in Figure \ref{fig:armsarrows}. The static version of Figure \ref{fig:armsarrows} shows the situation at the end of the simulation, with two field lines denoted by different symbols. The animated version of Figure \ref{fig:armsarrows} shows the situation every $100$~s.

\begin{figure*}[t!]
\centering
\includegraphics[scale=0.22]{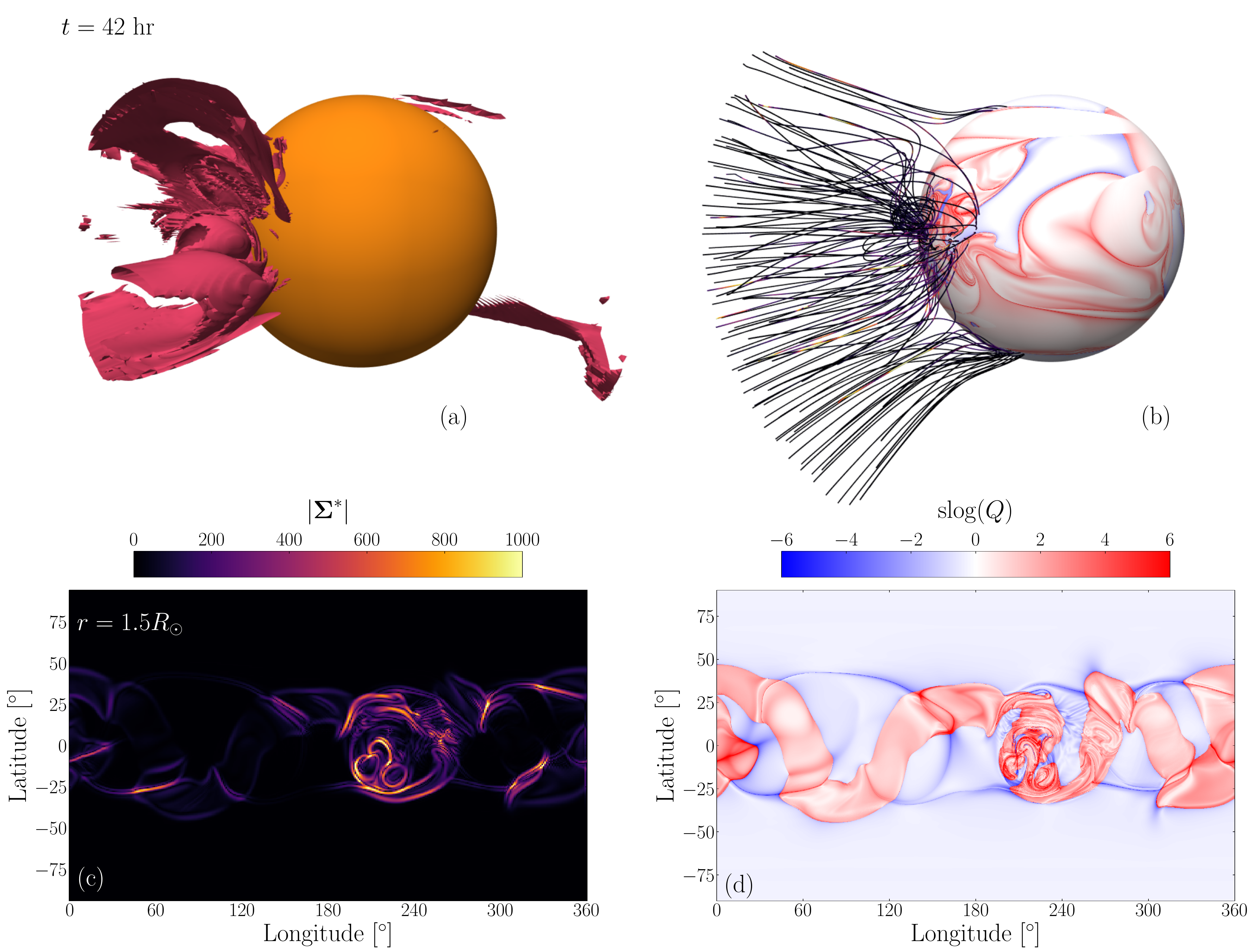}
\caption{An eruption occurring in a \texttt{DUMFRIC} simulation. (a) Isosurface of the slippage rate at $|\Sv^*|=500$ above the sun sphere. (b) A selection of field lines, colored by the slippage rate, above the sun sphere, colored by the squashing factor. (c) Slippage rate and (d) signed logarithm of the squashing factor in a slice at a constant radius $r=1.5 R_\odot$. Note that the slippage color scale is consistent across panels (a), (b) and (c). We denote 12:00 on 2017 August 24 as $t=0$, well before the start of the eruption and show the situation at $t=42$~hr. The animated version of this figure begins at $t=0$ and shows these quantities over six days, at times indicated.}
\label{fig:dumfricerupt}
\end{figure*}

It is also possible, at each time and field line intersection point, to calculate the local direction of the slippage rate $\Sv$. This is a local quantity, which does not require knowledge of long-range motion of the field line, or its footpoint. Arrows in both the static and animated versions of Figure \ref{fig:armsarrows} show the direction of the slippage rate (ignoring, again, the radial component). For the two field line intersections denoted by the triangle and pentagon, which fall into the ``O''-shape of the bottom right panel of Figure \ref{fig:arms3diag}, both the trajectory and the direction of slippage rate align almost perfectly, indicating that the field line is indeed slipping as described by the vector field $\Sv$. Note that in this simple case, the direction of $\Sv$ is approximately constant from $r=R_\odot$ to the height of Figure \ref{fig:armsarrows} at $r=1.2R_\odot$.

\subsection{Solar eruption (\texttt{DUMFRIC})}
\label{scn:dumfric}

The \texttt{DUMFRIC} code has recently been used to simulate the solar corona for a continuous period of 47 years \citep{aslanyan2024}. In this model, magnetic flux is injected into the corona at $r=R_\odot$ based on observational magnetograms. The simulation grid is initialized with a PFSS model. Thereafter, the magnetic field $\Bv=\nabla\times\Av$ is obtained by solving
\begin{align}
    \frac{\partial\Av}{\partial t} &= -\Ev     \label{eqn:induc} \\
    \Ev &= -\vv\times\Bv + \Nv, 
\end{align}
where $\Ev$ is the electric field and $\vv$ is the implied plasma velocity
\begin{equation}
    \vv = \frac{(\nabla\times\Bv)\times\Bv}{\nu |\Bv|^2} + v_w\left(\frac{r}{2.5R_\odot}\right)^{11.5}\ev_r.
\end{equation}
Here, $\nu$ is a frictional coefficient, $v_w=100$~km~s$^{-1}$ is the solar wind speed, and $\Nv$ is a hyperdiffusion term representing the macroscopic effects of small-scale turbulence. The simulation grid in $(r,\theta,\phi)$ has $61 \times 181 \times 361$ points, extending in a fully spherical shell from $R_\odot < r < 2.5R_\odot$.

\begin{figure}[h!]
\centering
\includegraphics[scale=0.25]{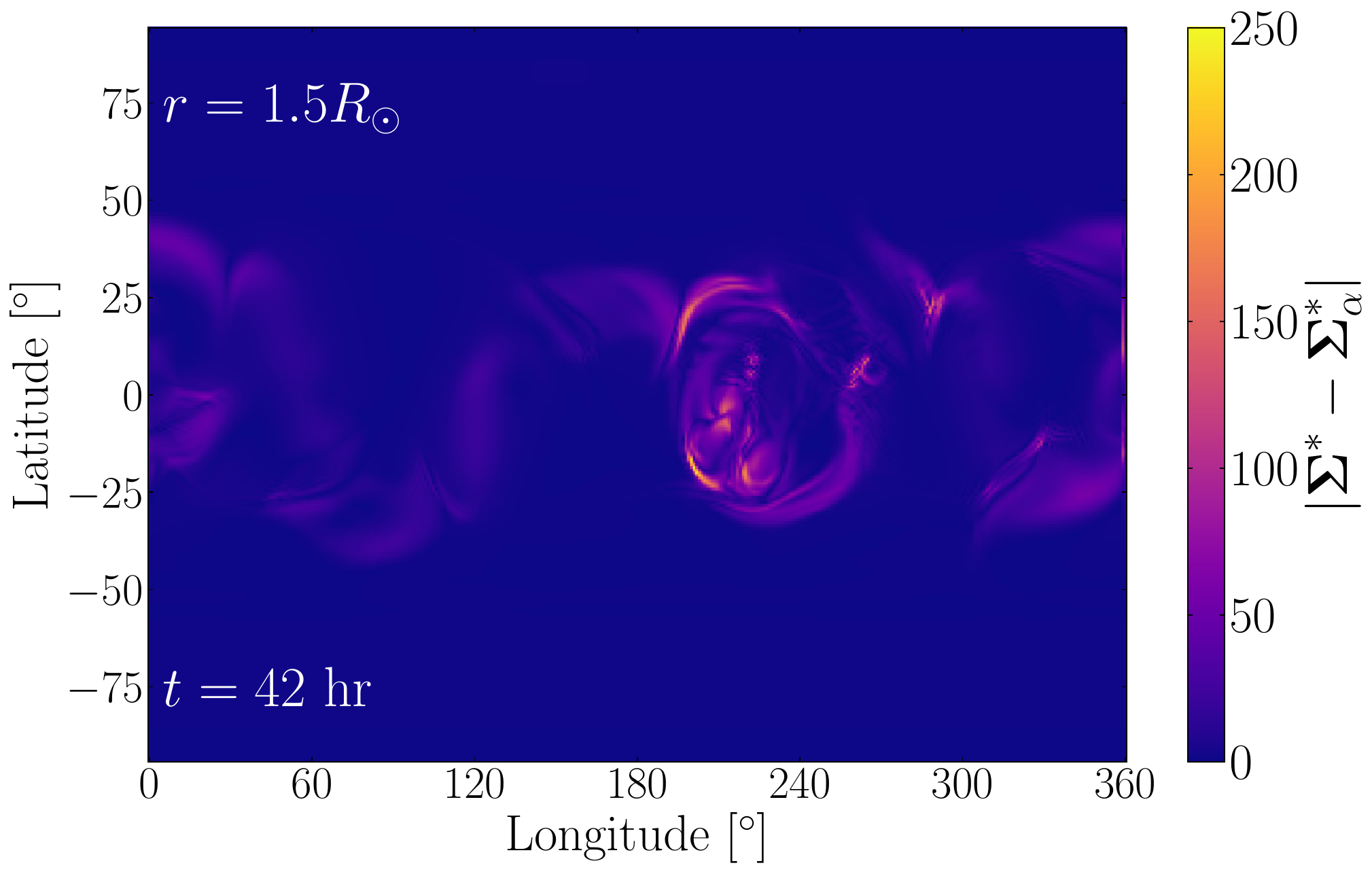}
\caption{During the \texttt{DUMFRIC} simulation, magnitude of the difference between the vector fields of full slippage, and slippage due to field aligned currents at $t=42$~hr. The animated version of this figure begins at $t=0$ and shows these quantities over six days, at times indicated.}
\label{fig:dumfricsigcomp}
\end{figure}

The system of units used by the \texttt{DUMFRIC} code is comparable to \texttt{ARMS}: $L=R_\odot$, current density is in units of Bi with the constant $4\pi$ replacing $\mu_0$, and $\eta$ is in units of ab$\Omega$~cm. Consequently, the scaled quantities output by \texttt{USlip} are related to their dimensional counterparts by Equations (\ref{eqn:armsscale1}) to (\ref{eqn:armsscale3}). As with the study in Section \ref{scn:arms}, the effective numerical resistivity is a constant to be determined.

\begin{figure}[h!]
\centering
\includegraphics[scale=0.2]{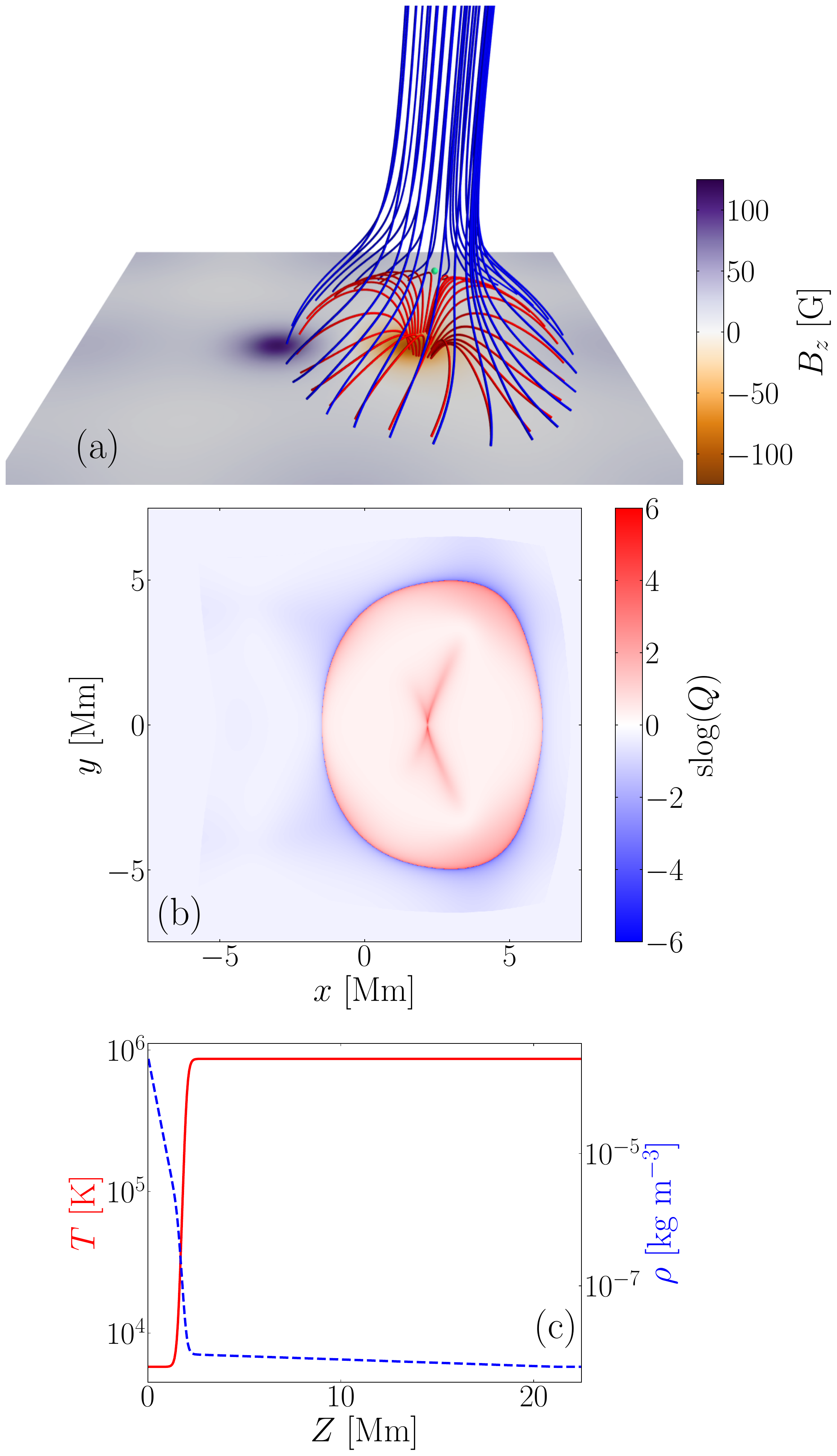}
\caption{Initial conditions of the Cartesian simulations with the \texttt{Lare3d} code outlined here. (a) Magnetic polarity at the photosphere above which are a sample of open (blue) and closed (red) magnetic field lines. The magnetic null is indicated by a green sphere. (b) Map of the signed logarithm of the squashing factor in the simulation domain. (c) Initial temperature (red, solid line) and density (blue, dashed line) profiles at heights above the photosphere as indicated, showing a sharp transition from the relatively cold, dense photosphere to the hot, tenuous corona.}
\label{fig:lare3dsetup}
\end{figure}

The long-term simulation of the corona was started at 12:00 on 1975 September 25. Active region 12672 emerged at $r=R_\odot$ at 12:00 on 2017 August 25, causing an eruption and subsequent relaxation of the coronal magnetic field. During this dynamic reorganization of the magnetic field, \texttt{USlip} identifies regions of field line slippage. Figure \ref{fig:dumfricerupt}(a) shows an isosurface of constant $|\Sv^*|=500$ which consists of a smaller, partially spherical structure surrounded by larger arcs intersecting the outer boundary. In the animated version of the Figure, which shows the situation every $6$~hr, a slippage ``bubble'' (visually reminiscent of bubble gum) expands outwards from where the active region emerges, suggesting that this is the mechanism by which non-ideal effects modify the field structure dynamically. A cut at a constant radius $r=1.5R_\odot$ is shown in Figure \ref{fig:dumfricerupt}(c), which displays the cross-section of this ``bubble''; it expands and contracts in the animated version. Some fine structure is seen to the South-West of the center.

Such a transient event is accompanied by strong twisting of the field lines and rapid variation in the squashing factor. Figure \ref{fig:dumfricerupt}(b) shows such twisted field lines colored by $|\Sv^*|$ above a sphere at $r=R_\odot$ colored by the squashing factor there. In the animated version of the figure, the twist propagates from the inner to the outer simulation boundary. The squashing factor is also shown at a constant radius $r=1.5R_\odot$ in Figure \ref{fig:dumfricerupt}(d). A number of regions of high, rapidly varying squashing factor with intermingled open and closed field lines correspond to the ``bubble'' of high slippage rate, with the fine structure of both diagnostics matching. This comparison of the slippage rate and squashing factor maps provides another clear example that while $Q$ indicates where the magnetic field is susceptible to reconnection, $\Sv$ indicates where reconnection is active at a given time \citep{StanishMacTaggart2026}.

\begin{figure*}[t!]
\centering
\includegraphics[scale=0.24]{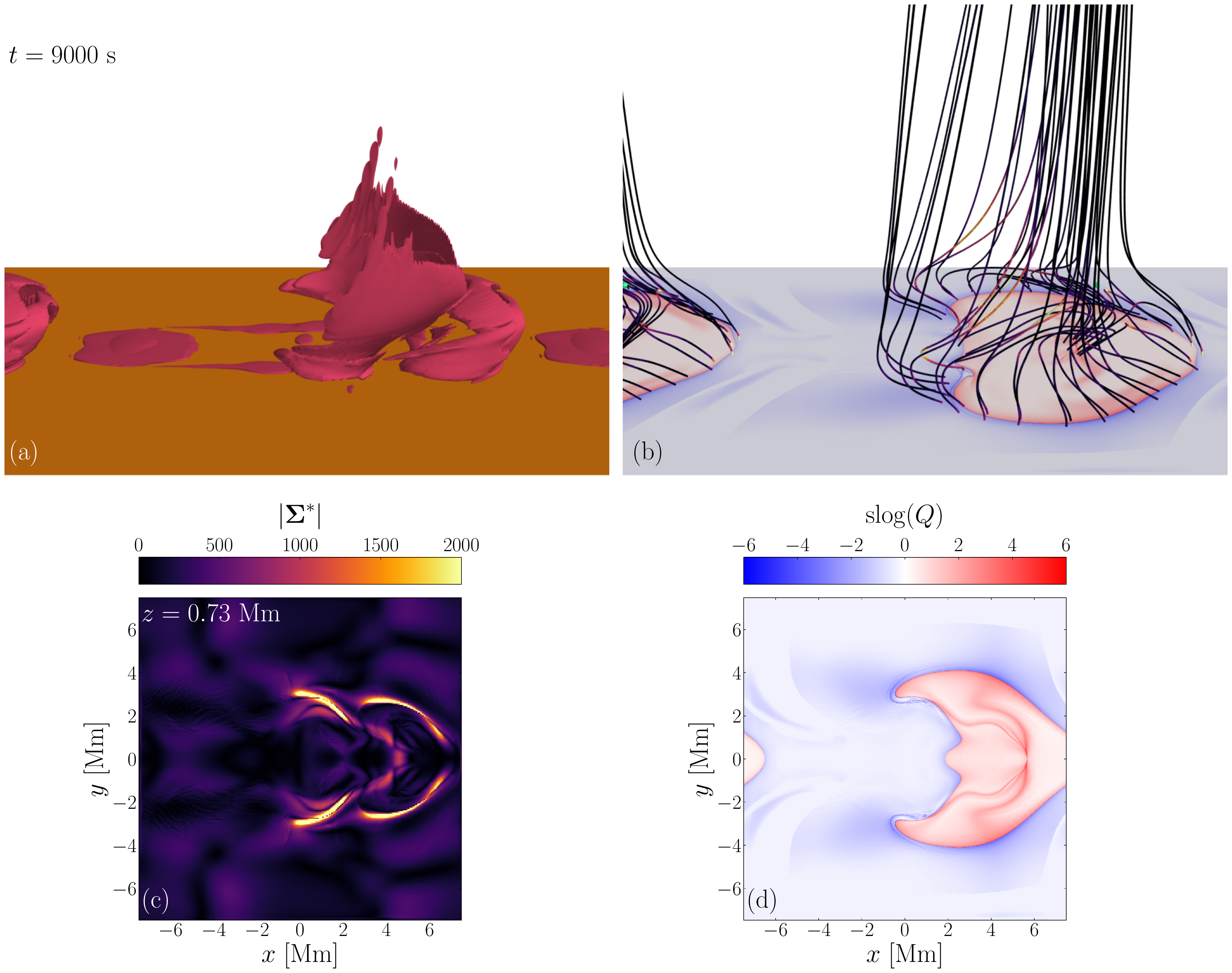}
\caption{Photospheric surface motion in the $+x$ direction (left to right on the page) leads to the motion and slippage of field lines. (a) Isosurface of the slippage rate at $|\Sv^*|=1000$ above the photosphere. (b) A selection of field lines, colored by the slippage rate, above the photosphere, colored by the squashing factor. (c) Slippage rate and (d) signed logarithm of the squashing factor in a slice at a constant height $z=0.73$~Mm. Note that the slippage color scale is consistent across panels (a), (b) and (c). Note that the 3-dimensional plots make use of the fact that the grid is periodic (and therefore infinitely repeating) in $x$. This is the situation at $t=9000$~s, while the animated version of this figure begins at $t=0$ and shows these quantities up to $t=20000$~s. A magnetic null, when it is identified, is indicated by a green sphere.}
\label{fig:lare3d}
\end{figure*}

Figure \ref{fig:dumfricsigcomp} displays a map of $|\Sv^*-\Sv_\alpha^*|$, similar to Figure \ref{fig:armssigcomp}, corresponding to the same time and location shown in Figure \ref{fig:dumfricerupt}(c). As with the example in Section \ref{scn:arms}, the cross-field gradients in $\alpha$ dominate the slippage rate magnitude. Although localized differences between $\Sv^*$ and $\Sv_\alpha^*$ are present, these remain small compared with the characteristic magnitude of $|\Sv^*|$, never exceeding approximately one fifth of its peak value. The neglected terms, therefore, contribute only a sub-dominant correction to the slippage rate. 

This example demonstrates that the theoretical and numerical framework laid out in this article is applicable to a range of models, remaining useful even when the full MHD equations are not solved. We anticipate that $\Sv$ will prove a useful diagnostic for locating regions of magnetic reconnection in studies where the topology is complicated and rapidly varying. The precise mechanisms of slippage and reconnection during eruptions, including the expanding non-ideal ``bubble'', identified in this model require further study.

\subsection{Magnetic geometry under photospheric motion (\texttt{Lare3d})}
\label{scn:lare3d}

To observe slippage rate on a smaller scale, we use the \texttt{Lare3d} MHD code \citep{lare3d} to simulate the interface between the photosphere and lower corona with a Cartesian grid. The simulation grid in $(x,y,z)$ measured $15\times 15 \times 22.5$ Mm, with $256^3$ grid points. A set of magnetic dipoles is used to initialize a region of parasitic polarity at the photosphere (located at $z=0$) leading to a region of closed field lines (defined as those which connect to the photosphere at both ends) embedded within the open field (defined here by field lines which connect from the photosphere out to $z>20$ Mm). This forms a ``dome'' of closed field lines as shown in Figure \ref{fig:lare3dsetup}(a). This geometry admits a magnetic null above the photosphere at the apex of the dome. The closed field region is initially approximately $10$ Mm in lateral size, as shown by the squashing factor map in Figure \ref{fig:lare3dsetup}(b). The initial background plasma temperature and density is intended to mimic the sharp transition region from the relatively cold photosphere to the hot corona. The temperature and density profiles, initially constant in $x$ and $y$, are shown in Figure \ref{fig:lare3dsetup}(c). The \texttt{Lare3d} code solves the resistive MHD equations under gravity, including heat conduction.

The domain is periodic in $x$ and $y$, allowing the system to be driven laterally, simulating photospheric flows. The simulations presented here are very similar to those described in \cite{pontin2024} -- wherein the overall dynamics are described in detail -- except that in that earlier study no chromospheric layer was included. To initiate the photospheric flow, at the layer of the lowest grid cells in $z$, we have imposed an additional force term composed of three Gaussian functions
\begin{equation}
    F_x = \sum_{i=1}^3 \rho \mathcal{T}_i \mathcal{W}_i \exp\left(-\frac{y^2}{w_i^2}\right),
\end{equation}
where $w_i$ is each corresponding width in $y$. Each of the time envelopes is sinusoidal,
\begin{align}
    \mathcal{T}_i &= \sin\left(\pi\frac{t-t_i}{T_i}\right), \quad t_i\leq t \leq t_i+T_i \\
    \mathcal{T}_i &= 0, \quad \mathrm{otherwise}
\end{align}
where $t_i$ is the individual start time and $T_i$ the period of each envelope. The envelope in $x$ is given by
\begin{align}
    \mathcal{W}_i &= \exp\left(\frac{-(x-L_i)^2}{\sigma_i^2}\right), \quad x\leq L_i \\
    \mathcal{W}_i &= 1, \quad\quad\quad\quad\quad\quad L_i\leq x \leq U_i \\
    \mathcal{W}_i &= \exp\left(\frac{-(x-U_i)^2}{\sigma_i^2}\right), \quad x \geq U_i
\end{align}
where $\sigma_i$ is the corresponding width, while the force is chosen to be constant between two limits $L_i$ and $U_i$, and to fall away as a Gaussian outside them.

The effect of this applied force is to move the plasma in the $+x$ direction, most intensely through the middle of the photosphere at $y=0$. After $7000$~s, the force is switched off and the system allowed to move freely under the resulting velocity. In the limit of ideal MHD, this would cause a lateral shift in field lines, which would remain frozen-in to the flow. On the other hand, the explicit and numerical resistivities cause the field lines to slip and, in many cases, to oscillate more intensely than under the strict influence of the fluid velocity. We have modified the \texttt{Lare3d} code to output the resulting surface velocity at sufficient frequency to allow us to track the footpoints of selected field lines.

Figure \ref{fig:lare3d}(a) shows an isosurface of constant $|\Sv^*|=1000$, which extends vertically upwards from the region of closed flux. The shape of this closed field region has been deformed from its initial state under the effect of the surface velocity, as shown in Figure \ref{fig:lare3d}(b). The dome of closed field lines and their surrounding open field lines is also shown, their footpoints having been advected by the surface flow. The field lines are colored by $|\Sv^*|$, with some field lines experiencing strong slippage close to, but not immediately at the photosphere. Figure \ref{fig:lare3d}(c) shows $|\Sv^*|$ in a slice at a distance $z=0.73$~Mm above the photosphere. Regions of high slippage form something resembling a ``wake'' around the moving dome of closed field lines. Figure \ref{fig:lare3d}(d) shows the squashing factor at this same height. One wake region forms just behind the leading open-closed boundary and the other at the trailing boundary.

The full dynamics are visible in the animated version of Figure \ref{fig:lare3d}, which shows the relative quantities every $1000$~s. The displayed field lines are first seen ``dragged'' by their footpoints and begin to slip close to the photosphere. Thereafter, the slippage extends further up the field lines and they begin to oscillate as they move in the $+x$ direction. The shape of the wake of slippage expands in time out from the closed field region.

\begin{figure}[h!]
\centering
\includegraphics[scale=0.36]{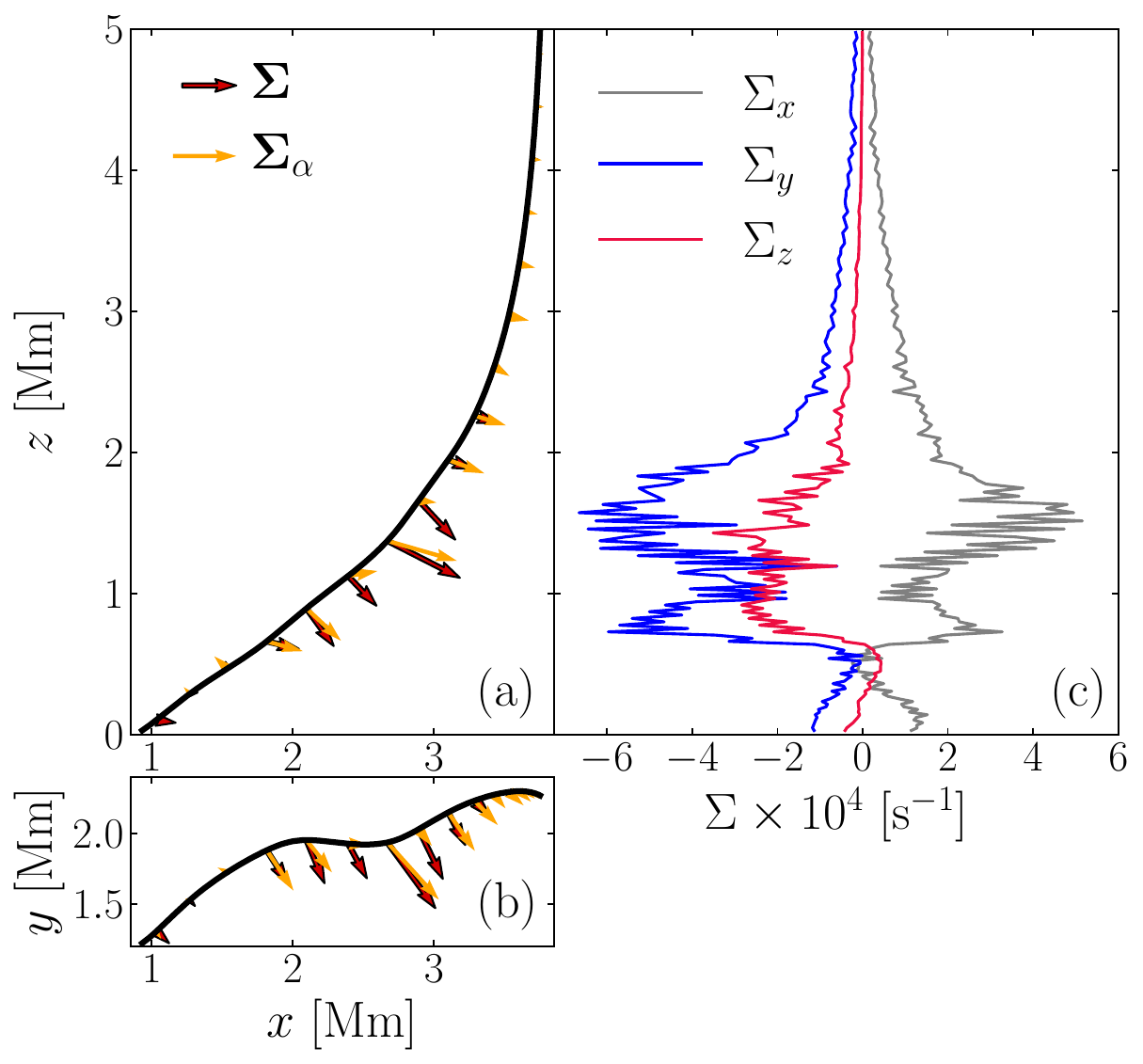}
\caption{At $t=9000$~s in the \texttt{Lare3d} simulation, a selected field line is projected onto the (a) $x-z$, (b) $x-y$ planes. The arrows indicate the magnitudes and directions of $\Sv$ and $\Sv_\alpha$ at points along the field line. (c) Variation of the components of $\Sv$ with height $z$. Note that the field line is shown up to a height of $z=5$~Mm, though the simulation grid extends above this. The field line slips most strongly between heights of $1\lesssim z\lesssim2$, directed towards $+x$ (parallel to the imposed flow), $-y$ (towards the center of the simulation region), and $-z$.}
\label{fig:slipfl}
\end{figure}

\texttt{Lare3d} normalizes quantities by a set of user-defined constants defined in SI units. We have chosen to set the length scale to $L=7.5$~Mm, and the resistivity to $\eta=11.9$~$\Omega$~m, while the vacuum permeability takes the usual value of $\mu_0=4\pi\times 10^{-7}$~N~A$^{-2}$. Note that the length scale has already been accounted for in the axes of the Figures shown here. These can be used with Equations (\ref{eqn:jstar}) to (\ref{eqn:svalphastar}) to obtain the absolute values of \texttt{USlip} outputs, in particular
\begin{align}
    \Sv &=\frac{\eta}{\mu_0 L^2}\Sv^* \\
    &=1.681\times 10^{-7}\Sv^*
\end{align}
where $\Sv$ is in units of s$^{-1}$.

We choose a particular open field line from Figure \ref{fig:lare3d}, at the rear (lower $x$) of the moving closed field region for further analysis. With the normalization constants above, we calculate $\Sv$ and $\Sv_\alpha$ dimensioned in units of s$^{-1}$ at points along the field line, as shown in Figure \ref{fig:slipfl}. We have cut off the field line at a height of $z=5$~Mm above the photosphere, so that its total length is $6.3$ Mm; the magnitudes of $\Sv$ and $\Sv_\alpha$ are low above there. This allows a cumulative slip ``velocity'' in Equation (\ref{eqn:sigmaintegral}) to be calculated by integration up along the field line for both measures of slippage rate. At the peak height shown here, we have
\begin{align}
    &\vv_{\mathrm{slip}} \;= (7.45,-11.2,-3.94)^T \times 10^{-4}\;\mathrm{Mm}\;\mathrm{s}^{-1}, \label{vslip vec} \\
    |&\vv_{\mathrm{slip}}| = 1.4\times 10^{-3}\;\mathrm{Mm}\;\mathrm{s}^{-1},\label{vslip mag} \\
    &\vv_{\alpha,\mathrm{slip}} \;= (6.82,-7.22,-1.56)^T \times 10^{-4}\;\mathrm{Mm}\;\mathrm{s}^{-1}, \\
    |&\vv_{\alpha,\mathrm{slip}}| = 1.0\times 10^{-3}\;\mathrm{Mm}\;\mathrm{s}^{-1}.\label{vslip alpha mag}
\end{align}
The quantities in \eqref{vslip vec} to \eqref{vslip alpha mag}, together with an inspection of the vectors in Figures \ref{fig:slipfl} (a) and (b), show that $\Sv_\alpha$ dominates other contributions to $\Sv$ on this field line, being ultimately responsible for both the strength and direction of the field line's slippage.

We can make further use of $\vv_{\rm slip}$ to estimate the distance of field line slippage (always relative to ideal motion), at least over suitably small time spans. Assuming that $\vv_{\rm slip}$ does not vary substantially over a time period $\Delta t$, then the slippage distance relative to ideal motion may be estimated as
\begin{equation}
    \rv_{\rm slip} \approx\vv_{\rm slip}\Delta t.
\end{equation}
In the present example, if $\Delta t=10^3$ s, the total slip of the field line at its top (here 5 Mm above the photosphere) would be 1.4 Mm relative to its ideal motion. We stress that $\rv_{\rm slip}$ is a simple estimate. However, it provides useful quantitative insight that would otherwise be computationally impractical to obtain.

\section{Conclusion}
\label{scn:conclusion}

We have presented \texttt{USlip}, a numerical tool for calculating the slippage rate of magnetic field lines and diagnosing non-ideal evolution in coronal magnetic fields. The method extends the \texttt{UFiT} framework to compute scaled versions of both the full slippage rate, $\Sv^*$, and its force-free approximation, $\Sv^*_\alpha$, providing spatially resolved and physically grounded diagnostics directly from magnetic field data inputs. 

The theoretical formulation underlying the slippage rate, based on the non-ideal contributions to the induction equation, has been summarized, together with its interpretation as an instantaneous and local measure of the deviation from ideal frozen-in motion. In contrast to geometrical diagnostics such as the squashing factor, which identify regions favorable for reconnection, the slippage rate directly quantifies where reconnection is actively occurring and provides the local direction and magnitude of changes in field line connectivity.

The numerical implementation has been validated against an analytical magnetized Z-pinch configuration (sometimes referred to as a ``screw pinch'', an approximation to a large major radius tokamak), where \texttt{USlip} accurately reproduces both the magnitude and direction of the slippage rate across coordinate systems, with controlled errors arising primarily at grid boundaries. This confirms the consistency of the discretization procedure and its suitability for practical applications to structured simulation grids. 

Applications to three different numerical simulations (based on the \texttt{ARMS}, \texttt{DUMFRIC} and \texttt{Lare3d} codes) demonstrate the capability of \texttt{USlip} to analyze complex, time-dependent magnetic configurations. 

In the magnetohydrodynamic (MHD) \texttt{ARMS} simulation, supergranular driving generates twisted bundles of magnetic field lines and spatially localized regions of enhanced slippage rate. These regions form coherent tubular structures aligned with the underlying flows, with their geometry reflecting the topology of the field: radial tubes in regions of nearly radial field, and curved, filamentary intersections where strongly inclined field lines are sampled on spherical shells. The slippage rate further reveals the direction of field line motion, capturing both rotational patterns driven by surface flows and the deformation of field lines at specific segments along their length.

In the magnetofrictional \texttt{DUMFRIC} simulation, \texttt{USlip} identifies the emergence of a spatially extended ``bubble'' of high slippage rate associated with an eruptive event. This structure expands and evolves dynamically. Its slippage rate signature refines that of its squashing factor signature, isolating the regions where non-ideal behavior is active rather than where it is potentially active.

At smaller scales, the MHD \texttt{Lare3d} simulation shows how photospheric driving produces a wake-like distribution of high slippage rate surrounding a moving dome of closed flux. The slippage rate initially develops near the photosphere and subsequently propagates along field lines, consistent with the progressive breakdown of ideal evolution. 

Across all examples, \texttt{USlip} not only localizes regions of strong non-ideal behavior but also provides a consistent description of the direction and magnitude of the field line slippage rate. The agreement between the inferred slippage direction and the observed evolution of field line trajectories (\textit{e.g.} Figure \ref{fig:armsarrows}) further supports the interpretation of $\Sv$ as a physically meaningful diagnostic of connectivity change. Calculations also show that $\Sv_\alpha$ dominates other contributions to $\Sv$ for the examples presented, as is to be expected in coronal applications.

The results demonstrate that \texttt{USlip} is a robust and versatile tool for the analysis of three-dimensional magnetic reconnection in complex systems. Its ability to operate on a wide range of inputs, including full MHD simulations, magnetofrictional models and analytically constructed fields, makes it well suited for both detailed analysis and comparative studies across different modelling approaches. It also shows promise for future simulations and observational studies where the 3
three-dimensional magnetic field is known. Regions of high scaled slippage rate $\Sv^*$ would indicate changes in magnetic connectivity, reconnection and other dynamic events. Where the resistivity is known, the absolute measure of slippage rate $\Sv$ may be calculated to compare directly to observations, with implications for turbulence, the release of solar wind and coronal heating.

It may furthermore be desirable to invert the problem presented here. Suppose that high resolution, time dependent maps of the magnetic field were obtained from observations. Simultaneously, particular loops of plasma could be identified and tracked in time based on their density or temperature contrast with the background plasma. Thus, by comparing the slippage rate and the true slippage of the plasma relative to field lines, the approximate resistivity of the plasma could be inferred.

\section*{Acknowledgements}

This work used the DiRAC Data Intensive service (CSD3) at the University of Cambridge, managed by the University of Cambridge University Information Services on behalf of the STFC DiRAC HPC Facility (www.dirac.ac.uk). The DiRAC component of CSD3 at Cambridge was funded by BEIS, UKRI and STFC capital funding and STFC operations grants. DiRAC is part of the UKRI Digital Research Infrastructure. Some of the results were obtained using the ARCHIE-WeSt High Performance Computer (www.archie-west.ac.uk) based at the University of Strathclyde. VA and DM acknowledge support from a Science and Technologies Facilities Council (STFC) grant (ST/Y001672/1). DM further acknowledges support from a Leverhulme Trust grant (RPG-2023-182). KAM acknowledges support from Science and Technologies Facilities Council (STFC) grant ST/W001098/1. PW acknowledges support from an STFC small grant (UKRI14137) and a Leverhulme Trust project grant (RPG-2023-288). RBS acknowledges support from NASA under ROSES HSR grant number 80NSSC26K0205. SKA acknowledges support from a NASA LWS grant and an NSF collaboration grant to the University of Michigan.

\appendix
\section{Comparing the diagnostics of GMR and the slippage rate}
\label{app:gmr}

A key diagnostic in the theory of general magnetic reconnection \citep[GMR;][]{Schindler1988,Hesse1988} is  the field-line voltage,
\begin{equation}
    \Xi = \int_L\frac{\Rv\cdot\Bv}{|\Bv|}\,{\rm d}\ell,
\end{equation}
where $L$ is a parameterization of the field line  (typically through a non-ideal region in the context of GMR theory). 

It is common to calculate $\Xi$ in simulations as a measure of reconnection. Whilst on many occasions this measure will diagnose reconnection, it is not, by itself, sufficient to imply changes in magnetic connectivity. To illustrate this, consider a linear force-free field satisfying
\begin{equation}
\nabla\times\Bv=\alpha_0\Bv,
\end{equation}
for constant $\alpha_0$, together with $\uv=\boldsymbol{0}$ and constant resistivity $\eta$. In this case,
\begin{equation}
\Rv
=\eta\jv
=\frac{\eta\alpha_0}{\mu_0}\Bv,
\end{equation}
and so
\begin{equation}
\nabla\times\Rv
=
\frac{\eta\alpha_0^2}{\mu_0}\Bv.
\end{equation}
Hence, $(\nabla\times\Rv)_\perp=\boldsymbol{0}$ and 
\begin{equation}
\mathbf{\Sv}=\mathbf{0},
\end{equation}
everywhere. At the same time, $\Xi$ is generally non-zero.

Thus, in this example, $\Xi$ measures magnetic diffusion along field lines but does not imply changes in field-line connectivity, since the slippage rate is identically zero. This illustrates that the field-line voltage and the slippage rate may diagnose different aspects of non-ideal evolution: a non-zero field-line voltage may be present without field-line slippage, whereas the slippage rate directly measures departures from frozen-in evolution.

\section{Slippage rate under resistivity}\label{appendix:slip_rate}

For completeness, we display the full expansion of the slippage rate $\Sv$ with $\Rv=\eta\jv$ and the decomposition identified in equations (\ref{eq:cur_dec1}) to (\ref{eq:alpha}). Following \cite{mactaggart2025}, we have
\begin{equation}
    \Sv = -\frac{1}{\mu_0|\Bv|}(\Cv_1 + \eta\Cv_2),
\end{equation}
where 
\begin{equation}
    \Cv_1 = \alpha\nabla\eta\times\Bv - \lambda(\nabla\eta\cdot\Bv)\ev_F,
\end{equation}
and
\begin{equation}
    \Cv_2 = (\lambda\omega_1-\nabla\lambda\cdot\Bv)\ev_F + \lambda(\alpha+\omega_2)\Bv_{f\perp} +\nabla\alpha\times\Bv,
\end{equation}
with 
\begin{equation}
    \omega_1=\nabla\times\Bv_{f\perp}\cdot\ev_F, \quad \omega_2 = \frac{(\nabla\times\Bv_{f\perp})\cdot\Bv_{f\perp}}{|\Bv|^2}. 
\end{equation}
In this work, since our focus is on large-scale coronal simulations, we have only considered constant resistivity, so the terms that constitute $\Cv_1$ are zero. The rest of the terms are based on the geometric scalars $\lambda$ and $\alpha$, with the last term in $\Cv_2$ contributing the ``force-free'' slippage rate $\Sv_\alpha$.

\section{Squashing Factor Calculation in Cylindrical Coordinates}
\label{scn:ufitcyl}

In cylindrical coordinates $(r,\phi,z)$, the position along a field line, parameterized by the distance $l$ along said field line, is given by 
\begin{align}
    \frac{ {\rm d} r}{ {\rm d} \ell} & = \frac{B_r}{|{\bf B}|},\\
    \frac{ {\rm d} \phi}{ {\rm d} \ell} & = \frac{1}{r} \frac{B_\phi}{|{\bf B}|},  \\
    \frac{ {\rm d} z}{ {\rm d} \ell} & = \frac{B_z}{|{\bf B}|}.
\end{align}
We have extended the theoretical and computational formalism for the squashing factor $Q$ outlined in \cite{ufitarticle} to cylindrical coordinates. Following the approach outlined there, a pair of tangent vectors $(\Uv, \Vv)$ are transported along the magnetic field according to the Lie transport
\begin{equation}
(\Bv \cdot \nabla) \Uv = (\Uv \cdot \nabla) \Bv,
\end{equation}
where $\nabla\beta \equiv \ev_i\, g^{ij} \partial_{x_j} \beta$ is the gradient of an arbitrary scalar field $\beta$, with $g^{ij}$ representing the inverse of the coordinate metric. Decomposed into its components, this expression takes the form
\begin{equation}
|\Bv| \frac{d}{d\ell} U_k \equiv (\Bv \cdot \nabla) U_k = (\Uv \cdot \nabla) B_k + (B_i U_j - U_i B_j) \Gamma^k_{ji}, 
\end{equation}
where $\Gamma^k_{ji} = g^{km} \ev_m \cdot \partial_{x_j} \ev_i$ are the Christoffel symbols that describe the local variation of the coordinate bases.

In cylindrical coordinates, the gradient of a scalar field $\beta$ is given by
\begin{equation}
    \nabla \beta = \left(\ev_r \partial_r + \frac{1}{r}\, \ev_\phi \partial_\phi+ \ev_z \partial_z \right) \beta \label{eqn:gradcyl}
\end{equation}
and the Christoffel symbols admit only two non-trivial entries, arising from the fact that $\partial_\phi \ev_r = \ev_\phi$ and $\partial_\phi \ev_\phi = -\ev_r$.
Of these two entries, only the off-diagonal term $\Gamma^{\phi}_{\phi r}$ contributes to the final expression (since $B_r U_r - U_r B_r = 0$).
Consequently, the complete set of transport equations in cylindrical coordinates is given by:
\begin{align}
    |\Bv|\frac{d}{d\ell} U_r & = (\Uv \cdot \nabla) B_r \\
    |\Bv|\frac{d}{d\ell} U_\phi & = (\Uv \cdot \nabla) B_\phi + \frac{1}{r} (U_\phi B_r - B_\phi U_r) \\
    |\Bv|\frac{d}{d\ell} U_z & = (\Uv \cdot \nabla) B_z.
\end{align}
Given solutions to these equations for initialized values of $\Uv$ and $\Vv$, which are taken to be initially orthonormal and orthogonal to $\Bv$, the squashing factor $Q$ is found in the same manner as for the other coordinate systems.

\section{Numerical derivatives}
\label{scn:derivatives}

Calculation of $\Sv$ requires the computation of the curl of a pair of vector fields, while $\Sv_\alpha$ additionally requires the gradient of a scalar field. These are defined in Cartesian, spherical, and cylindrical coordinates through their usual formulae, \textit{e.g.} the gradient in cylindrical coordinates is given by Equation (\ref{eqn:gradcyl}).

Suppose that in a 3-dimensional system of coordinates $(\rho,\sigma,\tau)$ some quantity $A$ is defined on a grid. On such a grid, for point $(\rho_i,\sigma_j,\tau_k)$, we assume that there exists at least one neighbor out of $(\rho_{i+1},\sigma_j,\tau_k),(\rho_{i-1},\sigma_j,\tau_k)$, and similarly for the other coordinates. In some cases, such as the \texttt{DUMFRIC} code, this is true of the entire simulation grid, while for the adaptive mesh of \texttt{ARMS} this is true for subsets of the grid at constant refinement. For a derivative with respect to coordinate $\rho$ at the point $(\rho_i,\sigma_j,\tau_k)$, we seek an ``upper'' point $\rho_\uparrow$ at $(\rho_{i+1},\sigma_j,\tau_k)$ if it exists or the point itself, and a ``lower'' point at $(\rho_{i-1},\sigma_j,\tau_k)$ or the point itself, with the derivative taking the usual form
\begin{equation}
   \left. \frac{\partial A}{\partial \rho}\right|_{\rho_i,\sigma_j,\tau_k} \approx \frac{A_\uparrow-A_\downarrow}{\rho_\uparrow-\rho_\downarrow},
\end{equation}
and similarly for the other coordinates $\sigma,\tau$. This reduces to a first- or second-order method respectively.

Outside of Cartesian coordinates, the gradient and curl require the division by scale factors, \textit{e.g.} $1/r$ in Equation (\ref{eqn:gradcyl}). We find that, for the desired mixture of speed and accuracy, the functions within the respective scale factors are evaluated at the midpoint of the upper and lower point, as
\begin{equation}
    f(\rho) \left. \frac{\partial A}{\partial \rho}\right|_{\rho_i,\sigma_j,\tau_k} \approx f\left(\frac{\rho_\uparrow+\rho_\downarrow}{2}\right)\frac{A_\uparrow-A_\downarrow}{\rho_\uparrow-\rho_\downarrow},
\end{equation}
and similarly for the other coordinates. As stated above, the second order derivatives in the definition of $\Sv$ and $\Sv_\alpha$ are calculated by two successive passes over the simulation grid.

\bibliography{bibliog}{}
\bibliographystyle{aasjournal}

\end{document}